\documentclass[noshowpacs,preprintnumbers,amsmath,amssymb]{revtex4}

\usepackage{graphicx}
\usepackage{bm}
\usepackage{dcolumn}
\usepackage{color}
\usepackage{pst-plot}

\newcommand{\be}{\begin{equation}}
\newcommand{\ee}{\end{equation}}
\newcommand{\bea}{\begin{eqnarray}}
\newcommand{\eea}{\end{eqnarray}}
\newcommand{\non}{\nonumber}

\begin{document}

\title{Step-like discontinuities in Bose-Einstein condensates and Hawking radiation: dispersion effects}

\author{Carlos Mayoral}
\email{carlos.mayoral@uv.es}
\affiliation{Departamento de F\'isica Te\'orica and IFIC, \\ Universidad de Valencia-CSIC, C.\ Dr.\ Moliner 50, 46100 Burjassot, Spain}

\author{Alessandro Fabbri}
\email{afabbri@ific.uv.es}
\affiliation{Departamento de F\'isica Te\'orica and IFIC,\\ Universidad de Valencia-CSIC, C.\ Dr.\ Moliner 50, 46100 Burjassot, Spain\\}
\affiliation{APC (Astroparticules et Cosmologie),   \\10 Rue A.\ Domon et L.\ Duquet, 75205 Paris Cedex 13, France}

\author{Massimiliano Rinaldi}
\email{massimiliano.rinaldi@unige.ch}
\affiliation{D\'epartment de Physique Th\'eorique, \\ Universit\'e de Gen\`eve, \\24 quai E.\ Ansermet  CH--1211, Gen\`eve 4, Switzerland\\}

\date{\today}

\begin{abstract}
\noindent
In this paper we extend the hydrodynamic results of \cite{nostro} and study, analytically, the propagation of Bogoliubov phonons on top of Bose-Einstein condensates with step-like discontinuities by taking into account dispersion effects. We focus on the Hawking signal in the density-density correlations in the formation of acoustic black hole-like configurations.
%Max's abstract
%In this paper we analyze the correlations between Hawking partners in condensates with step-like discontinuities in the speed of sound. In particular,  we study analytically the propagation of Bogoliubov phonons on top of Bose-Einstein condensates with step-like discontinuities by taking into account also dispersion effects, thus extending the results of \cite{nostro}. We also focus on the Hawking signal in the density-density correlations in the formation of acoustic black hole-like configurations, and we obtain their analytic approximations. Finally, we show that these are in good agreement with numerical calculations, and that the temporal formation of the step amplifies the correlation signal.
\end{abstract}


\maketitle

\section{Introduction}\label{intro}

\noindent The study of analog models of gravity in condensed matter systems \cite{unruh,blv} has motivated the investigation of quantum effects in gravity, in particular Hawking radiation from black holes \cite{hawking}, in the presence of modified dispersion relations (see \cite{unruh2,UJC,bffp}). Modified dispersion relations at high frequency have also been considered in many papers in cosmology (see e.g. \cite{cosmo}), but also in the context of the Unruh effect \cite{maxunruh}, which is closely related to the Hawking emission from a black hole. On a more formal level, issues related to quantum field renormalization in the presence of dispersion were investigated in \cite{maxdisp}. Among the many systems proposed to create black hole-like configurations, e.g. superfluid liquid Helium \cite{volovik}, atomic Bose-Einstein condensates (BECs) \cite{gc}, surface waves in water tanks \cite{rousseaux}, degenerate Fermi gases \cite{fermi}, slow light in moving media \cite{light}, traveling refractive index interfaces in non linear optical media \cite{optica}, BECs, characterized by superluminal dispersion relations, appear to be quite attractive from the experimental point of view \cite{blv2}. In this context, recently an alternative measure of the Hawking effect was proposed in terms of non local density correlations \cite{fteo} for the Hawking quanta and their partners situated on opposite sides with respect to the acoustic horizon. The calculations were performed using the gravitational analogy, which corresponds to the hydrodynamic approximation of the theory. This proposal was validated with numerical simulations within the microscopic theory \cite{fnum}, indicating that the Hawking signal in the correlations is indeed robust. Subsequent investigations were performed in \cite{recatipavloff} (where analytical approximations based on step-like discontinuities were considered) and \cite{cp1} using stationary configurations.

In this paper we extend the hydrodynamical analysis in \cite{nostro} and consider in particular the effects of the temporal formation of acoustic black hole-like configurations, as in \cite{fteo} and \cite{fnum}, including dispersion effects. Our analytical analysis is based on step-like discontinuities in the speed of sound and thus extends the stationary results in \cite{recatipavloff}. We mention that step-like configurations in BECs were also considered in \cite{barcelogaraydopio,jp,barcelogaray,wswav,ftemp}.

The plan of the paper is the following: in section II we briefly describe the model used and the basic equations, while in sections III and IV we analyze thoroughly the stationary case (spatial step-like discontinuities) and the homogeneous one (temporal step-like discontinuities). By combining the results of these two sections, in section V we discuss the main Hawking signal in correlations for the formation of acoustic black hole-like configurations and in section VI we end with comparisons with the hydrodynamical results in \cite{fteo}.

\section{\bf The model and its basic equations}\label{sec2}

\noindent We start with the basic equations  for a Bose gas in the dilute gas approximation described by a field operator $\hat \Psi$ \cite{ps}-\cite{castin}. The equal-time commutator is
\begin{equation}\label{etc}
[\hat \Psi (t,\vec x), \hat \Psi^{\dagger}(t,\vec x')]=\delta^3(\vec x- \vec x')
\end{equation}
and the time-dependent Schr\"odinger equation is given by
\begin{equation}
i\hbar \partial_t \hat \Psi = \left(-\frac{\hbar^2}{2m}\vec \nabla^2 + V_{ext} + g\hat \Psi^{\dagger}\hat \Psi\right)\hat \Psi\ ,
\end{equation}
where $m$ is the mass of the atoms, $V_{ext}$ the external potential and $g$ the nonlinear atom-atom interaction  constant. By considering the mean-field expansion
\begin{equation}\label{mfexp} \hat \Psi \sim \Psi_0 (1 + \hat \phi)\ ,\end{equation}
with $\hat\phi$ a small perturbation.
 The macroscopic condensate is described by the classical wavefunction $\Psi_0$ which satisfies the Gross-Pitaevski equation
\begin{equation}\label{gp}
i\hbar\partial_t \Psi_0 = \left(-\frac{\hbar^2}{2m}\vec \nabla^2 + V_{ext} + g n \right)\Psi_0\ ,
\end{equation}
where $n=|\Psi_0|^2$ is the number density, and the linear perturbation $\hat \phi$ satisfies the Bogoliubov-de Gennes equation
\begin{equation}\label{bdg}
i\hbar  \partial_t   \hat \phi= - \left( \frac{\hbar^2}{2m}\vec \nabla^2 + \frac{\hbar^2}{m}\frac{\vec \nabla \Psi_0 }{\Psi_0} \vec \nabla\right)\hat\phi +mc^2 (\hat\phi + \hat\phi^{\dagger})\ ,
\end{equation}
where  $c=\sqrt{\frac{gn}{m}}$ is the speed of sound.

%\subsection{Our model}
To study analytically the solutions to (\ref{bdg}), along the lines of \cite{fnum}, we shall consider condensates of constant density $n$ and velocity (for simplicity along one dimension, say $x$). Non-trivial configurations are still possible, provided one varies the coupling constant $g$ (and therefore the speed of sound $c$) and the external potential but keep the sum $gn+ V_{ext}$ constant. In this way, the plane-wave function $\Psi_0=\sqrt{n}e^{ik_0x-iw_0t}$, where $v=\frac{\hbar k_0}{m}$ is the condensate velocity, is a solution of (\ref{gp}) everywhere.

%Two different representations will be considered for the fluctuations.
The non-hermitean operator $\hat \phi$ is expanded as
\begin{equation}\label{frep}
\hat\phi (t,x) =\sum_j \left[ \hat a_j \phi_j (t,x) + \hat a_j^{\dagger} \varphi_j^*(t,x) \right]\ ,\end{equation}
where $\hat a_j$ and $\hat a_j^{\dagger}$ are the phonon's annihilation and creation operators.
From (\ref{bdg}) and its hermitean conjugate, we see that the modes $\phi_j (t,x)$ and
$\varphi_j(t,x)$ satisfy the coupled differential equations
\bea\non
\left[ i(\partial_t + v\partial_x) + \frac{\xi c}{2} \partial_x^2 -\frac{c}{\xi} \right] \phi_j &=& \frac{c}{\xi}
\varphi_j\ , \\
\left[ -i (\partial_t + v\partial_x) + \frac{\xi c}{2}\partial_x^2 - \frac{c}{\xi}\right] \varphi_j &=&\frac{c}{\xi} \phi_j  \ ,\label{cde}
\eea
 where $\xi=\hbar/(mc)$ is the so-called healing length of the condensate. The normalizations are fixed, via integration of the equal-time commutator obtained from (\ref{etc}), namely
\begin{equation}\label{etcd} [\hat \phi (t,x), \hat\phi^{\dagger}(t,x')]=\frac{1}{n}\delta(x-x')\ ,\end{equation}  by
%\begin{equation}
%\label{nor}
%
%\begin{equation}
%\label{cde}
%\begin{array}{c}
%\left[ i\hbar (\partial_t + v\partial_x) + \frac{\hbar^2}{2m} \partial_x^2 -mc^2 \right] \phi_j = mc^2
%\varphi_j\ , \\  \left[ -i\hbar (\partial_t + v\partial_x) + \frac{\hbar^2}{2m}\partial_x^2 - mc^2\right] \varphi_j = mc^2 \phi_j  \ .
%\end{array}
%\end{equation}
%The normalizations are fixed, via integration of the equal time commutator (from  (\ref{etc}))
%\begin{equation}\label{etcd} [\hat \phi (t,x), \hat\phi^{\dagger}(t,x')]=\frac{1}{n}\delta(x-x')\ ,\end{equation}  by
\begin{equation}
\label{nor}
\int dx [\phi_j\phi_{j'}^* - \varphi_j^*\varphi_{j'}]=\frac{\delta_{jj'}}{\hbar n}\ .\end{equation}

We shall consider step-like discontinuities in the speed of sound $c$, which is the only non-trivial parameter in this formalism, and impose the appropriate boundary conditions for the modes that are solutions to Eqs.\ (\ref{cde}). A similar analysis was carried out in the hydrodynamic limit $\xi \rightarrow 0$ in the work \cite{nostro}, by using the more appropriate density phase representation \begin{equation}\label{dph}
\hat\phi = \frac{\hat n^1}{2n} + i\frac{\hat\theta^1}{\hbar}\ . \end{equation}.

\section{Step-like spatial discontinuities (stationary case)}\label{sec3}

\noindent In this section we study dispersion effects for the case of spatial step-like discontinuities. We treat subsonic configurations in subsection \ref{subconf}, thus extending the hydrodynamic analysis of \cite{nostro}, and subsonic-supersonic ones in subsection \ref{supconf}.  This case is particularly interesting in view of our application to study the main Hawking signal in correlations from acoustic black holes, along the lines of \cite{fnum}.

\subsection{Subsonic configurations}\label{subconf}

We consider a surface (that we put for simplicity at $x=0$) separating two semi-infinite homogeneous condensates with different sound speeds: $c(x)=c_l\theta(-x) + c_r \theta(x)$. The velocity of the condensate is taken to be negative ($v<0$), so that the flow is from right to left. We assume that the condensate is everywhere subsonic, that is $|v|<c_{r(l)}$, and that $v$, $c_l$ and $c_r$ are time-independent.

To explicitly write down the decomposition of the field operator $\hat \phi$, we first need to study the propagation of the modes and construct the ``in'' and ``out'' basis.
To understand the details of modes propagation, we need to solve the equations (\ref{cde}) in the left and right homogeneous regions, and then impose the appropriate boundary conditions. These simply are the requirement that $\phi$ and $\varphi$, along with their first spatial derivatives, are continuous across the discontinuity at $x=0$.

\begin{figure}[htbp]
\begin{center}
%\resizebox{!}{8cm}{\input{uhidro.eps}}
\resizebox{!}{6cm}{\includegraphics{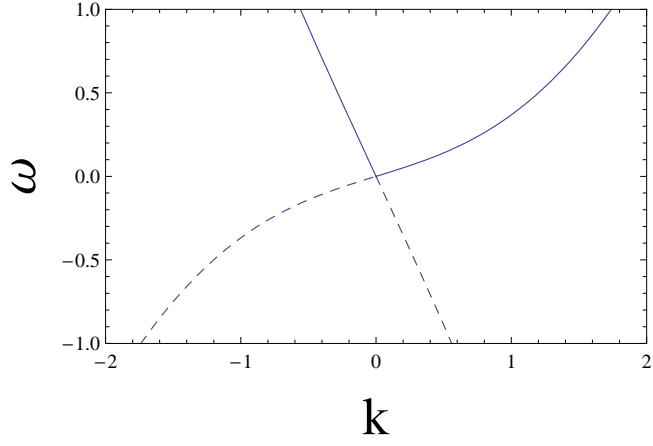}}
\caption{Dispersion relation for subsonic configurations. The solid (dashed) line corresponds to the positive (negative) norm branch: $\omega-vk=+(-)c\sqrt{k^2+\frac{\xi^2 k^4}{4}}$.}
\label{fig:sub_disp}
\end{center}
\end{figure}

We denote the modes solutions in each homogeneous region and corresponding to the fields $\phi$ and $\varphi$ as $De^{-iwt+ikx}$ and $Ee^{-iwt+ikx}$ respectively. The boundary conditions at the discontinuity, as we will see explicitly later, require us to work at fixed $\omega$. Therefore we write the modes as
\bea
\phi_{\omega}=D(\omega)e^{-iwt+ik(\omega)x}\ ,\qquad \varphi_{\omega}=E(\omega)e^{-iwt+ik(\omega)x}\ ,
\eea
so that the equations (\ref{cde}) simplify to
%
%The modes are, at fixed $\omega$, $\phi_w^{i}\equiv De^{-iwt+ik_{i}(w)x}$ and $\varphi_w^{i} \equiv E e^{-iwt+ik_{i}(w)x}$, where $k_{i}$ are the four roots of the dispersion relation ($\xi=\frac{\hbar}{mc}$)
%\begin{equation}\label{nrela}
%(w-vk)^2=c^2(k^2+ \frac{\xi^2 k^4}{4}).
%\end{equation}
%We shall call $k_v$ and $k_u$ the two real roots with negative and positive group velocity, respectively. The other two roots, being in the subsonic case, are complex. $k_d$ is the complex root with positive imaginary part, which represents a decaying mode for $x>0$ and a growing mode for $x<0$. Instead, $k_g$ is the complex root with negative imaginary part, that is a growing(decaying) mode for $x>0$($x<0$).\\

\bea\non\label{gupa}
\left[ (w-vk) - \frac{\xi c k^2}{2}  -\frac{c}{\xi} \right] D(\omega) &=& \frac{c}{\xi} E(\omega)\ , \\
\left[ - (w-vk) - \frac{\xi c k^2}{2} - \frac{c}{\xi}\right] E(\omega) &=& \frac{c}{\xi} D(\omega)  \ ,
\eea
while the normalization condition (\ref{nor}) ($j\equiv \omega$) gives
\bea\label{nodia}
 |D(\omega)|^2 - |E(\omega)|^2={1\over 2\pi \hbar n}\Big|\frac{dk}{dw}\Big|\ .
\eea
The combination of the two Eqs. (\ref{gupa}) gives the non linear dispersion relation
\begin{equation}\label{nrela}
(w-vk)^2=c^2\left(k^2+ \frac{\xi^2 k^4}{4}\right),
\end{equation}
plotted in Fig.\ \ref{fig:sub_disp}. At low momenta ($k\ll \frac{1}{\xi}$) we recover the linear relativistic dispersion, while at
large momenta ($k\gg \frac{1}{\xi}$) the nonlinear superluminal term becomes dominant. %In the dispersionless limit $\xi\to 0$ they reduce to their hydrodynamic values
%$k_v=\frac{w}{v-c}$ and $k_u=\frac{w}{v+c}$.
Moreover, inserting the relation between $D$ and $E$ from (\ref{gupa}) into (\ref{nodia}) we find the mode normalizations
\bea\label{eq:normdispersion}
D(\omega) &=&  \frac{\omega -v k+\frac{c\xi k^2}{2}}{\sqrt{4\pi \hbar n c\xi k^2\left| (\omega-vk) \left(\frac{dk}{d\omega}\right)^{-1} \right| }},\nonumber\\
E(\omega) &=& -\frac{\omega -v k-\frac{c\xi k^2}{2}}{\sqrt{4\pi \hbar nc\xi k^2\left| (\omega-vk) \left(\frac{dk}{d\omega}\right)^{-1} \right| }},
\eea
where $k=k(\omega)$ are the roots of the quartic equation (\ref{nrela}) at fixed $\omega$. Eq. (\ref{nrela}) admits, in the subsonic case, two real and two complex solutions. Regarding the real solutions,  we will call $k_v$ and $k_u$ the ones corresponding to negative and positive group velocity $v_g={d\omega\over dk}$ respectively. They admit a perturbative expansion in the dimensionless parameter $z\equiv {\xi \omega\over c}$, namely
\begin{eqnarray}\label{uv}
&&k_v=\frac{\omega}{v-c}\left(1+\frac{c^3z^2}{8(v-c)^3}+O(z^4)\right)\ ,\nonumber\\
&&k_u=\frac{\omega}{v+c}\left(1-\frac{c^3z^2}{8(v+c)^3}+O(z^4)\right)\ .
\end{eqnarray}
The other two solutions are complex conjugates. We call $k_d(k_g)$ the roots with positive(negative) imaginary part, which represent a decaying(growing) mode on the positive $x>0$ axis and a growing(decaying) mode in the negative ($x<0$) one. Such roots are non-perturbative in $\xi$ as they diverge in the hydrodynamic limit $\xi=0$, when Eq.\ (\ref{nrela}) becomes quadratic. However, they admit the expansions
\begin{equation}\label{decaying}
k_{d(g)}=\frac{\omega |v|}{c^2-v^2}\left[1-\frac{(c^2+v^2)c^4z^2}{4(c^2-v^2)^3}+O(z^4)\right]+(-)\frac{2i\sqrt{c^2-v^2}}{c\xi}\left[1+\frac{(c^2+2v^2)c^4z^2}{8(c^2-v^2)^3}+O(z^4)\right]\ .
\end{equation}
In what follows we do not need to specify the normalization coefficients for these modes, that we call generically $\frac{d_{\phi(\varphi)}}{\sqrt{4\pi n \hbar}}$ and $\frac{G_{\phi(\varphi)}}{\sqrt{4\pi n \hbar}}$ for the decaying and growing modes respectively, of the fields $\phi$ and $\varphi$.

In summary, the most general decompositions of $\phi$ and $\varphi$ in the left and right regions are given by
\begin{eqnarray}\label{decuno}
\phi_{\omega}^{l(r)} &=& e^{-i\omega t}\left[D_v^{l(r)}A_v^{l(r)}e^{ik_v^{l(r)}x}+D_u^{l(r)}A_u^{l(r)}e^{ik_u^{l(r)}x}+d_{\phi}^{l(r)}A_d^{l(r)}e^{ik_{g(d)}^{l(r)}x}+G_{\phi}^{l(r)}A_G^{l(r)}e^{ik_{d(g)}^{l(r)}x}\right]\ ,\\
  \varphi_{\omega}^{l(r)} &=& e^{-i\omega t}\left[E_v^{l(r)}A_v^{l(r)}e^{ik_v^{l(r)}x}+
  E_u^{l(r)}A_u^{l(r)}e^{ik_u^{l(r)}x}+d_{\varphi}^{l(r)}A_d^{l(r)}e^{ik_{g(d)}^{l(r)}x}+G_{\varphi}^{l(r)}A_G^{l(r)}e^{ik_{d(g)}^{l(r)}x}\right] \ . \label{decdue}
  \end{eqnarray}
The coefficients $A_{u,v,d,G}^{l,r}$ are the amplitudes of the modes, not to be confused with the normalizations coefficients. Indeed, the latter are determined uniquely by the commutation relations and the equations of motion, while the amplitudes depend on the particular choice of basis, as shown below. The matching conditions at $x=0$ to be imposed on Eqs.\ (\ref{cde}) are
\begin{equation}\label{matchingaa}
[\phi]=0,\, [\phi']=0,\, [\varphi]=0,\, [\varphi']=0,
\end{equation}
where [ ] indicates the variation across the jump. It is clear that these conditions require $\omega$ to be the same in the $l$ and $r$ regions.
Eqs. (\ref{matchingaa}) can be written in matrix form
\begin{equation}
%\label{eq:bogoc}\label{eq:bogoc}
W_l\left(
     \begin{array}{c}
       A_v^l \\
       A_u^l \\
       A_G^l \\
       A_d^l \\
 \end{array}
   \right)=W_r\left(
                \begin{array}{c}
                  A_v^r \\
       A_u^r \\
       A_d^r \\
       A_G^r \\
                \end{array}
              \right),
\end{equation}
where
\begin{equation}
\label{eq:wl}
W_l=\left(
     \begin{array}{cccc}
       D_v^l & D_u^l & G_{\phi}^l & d_{\phi}^l\\
       ik_v^lD_v^l & ik_u^lD_u^l & ik_g^lG_{\phi}^l & ik_d^ld_{\phi}^l \\
       E_v^l & E_u^l & G_{\varphi}^l & d_{\varphi}^l \\
       ik_v^lE_v^l & ik_u^lE_u^l & ik_g^lG_{\varphi}^l & ik_d^ld_{\varphi}^l  \\
\end{array}\right)
\end{equation}
and
\begin{equation}
\label{eq:wr}
W_r=\left(
     \begin{array}{cccc}
       D_v^r & D_u^r & d_{\phi}^r & G_{\phi}^r\\
       ik_v^rD_v^r & ik_u^rD_u^r &   ik_d^rd_{\phi}^r&ik_g^rG_{\phi}^r \\
       E_v^r & E_u^r & d_{\varphi}^r &  G_{\varphi}^r\\
       ik_v^rE_v^r & ik_u^rE_u^r & ik_d^rd_{\varphi}^r  &  ik_g^rG_{\varphi}^r\\
\end{array}\right).
\end{equation}
Multiplying both sides by $W_l^{-1}$ we have
  \begin{equation}
%\label{eq:matchingtc} no need for this label
     \left( \begin{array}{c}
       A_v^l \\
       A_u^l \\
       A_G^l \\
       A_d^l \\
     \end{array} \right)
   =M_{scatt} \left(
                \begin{array}{c}
                             A_v^r \\
       A_u^r \\
       A_d^r \\
       A_G^r \\
                \end{array}
              \right) \ .
\end{equation}
The $4\times 4$ matrix $M_{scatt}\equiv W_l^{-1} W_r$ encodes all non-trivial scattering effects due to the matching conditions (\ref{matchingaa}). The form of $M_{scatt}$ is much more involved than that found in the hydrodynamic limit in \cite{nostro}. 

To construct $M_{scatt}$ we have used the general decompositions (\ref{decuno}) and (\ref{decdue}). Not all modes, however, are physically meaningful. The validity of the mean-field approximation (\ref{mfexp}) implies that only spatially bounded modes have to be taken into account. This means that the amplitudes of the growing modes (that diverge exponentially in the $l$ or $r$ regions) must be set to zero. There are no constraints, instead, for the amplitudes of the decaying modes. Indeed, as we will see explicitly in the construction of the ``in'' and ``out'' modes basis that follows, by taking into account the ($l$ and $r$) decaying modes  we have each time four amplitudes which are uniquely determined by our four matching equations.
The physical meaning of the decaying modes is to ``dress'' the ``in'' and ``out'' modes basis, and this affects the calculation of local observables (this discussion follows that of \cite{Macher:2009tw}).

\begin{figure}[htbp]
\begin{center}
\resizebox{!}{12cm}{\input{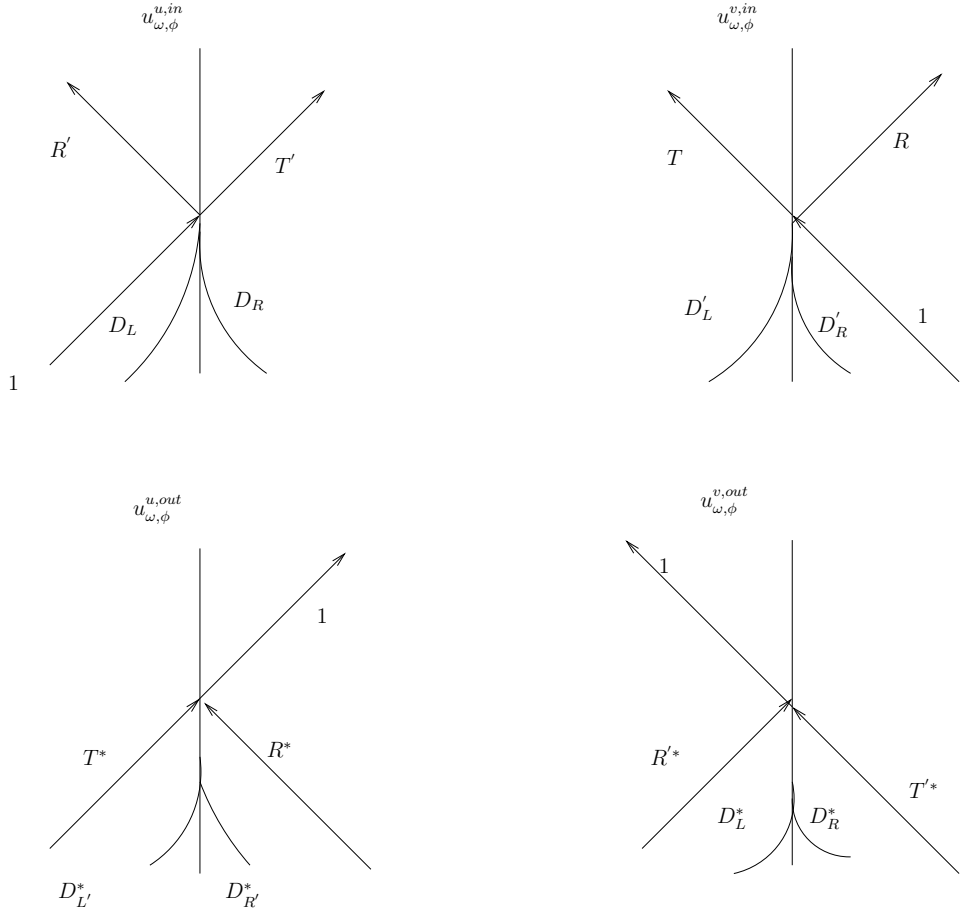}}
\caption{``in'' and ``out''  modes for spatial step-like discontinuities between homogeneous subsonic regions. We display the propagating modes (straight lines) and the decaying modes (curved lines), along with their amplitudes.}
\label{fig:disp_sub_in_out}
\end{center}
\end{figure}

We now proceed to construct the ``in'' and ``out'' modes basis for the case $v=0$ in a perturbative expansion up to $O(z^2)$. This case can also be treated exactly, as shown in the Appendix A. The perturbative construction of the ``in'' modes for the more complicated case $v\neq 0$ is given in Appendix B. To appreciate similarities and differences with respect to the hydrodynamical case treated in \cite{nostro}, let us construct perturbatively the ``in'' and ``out''  modes basis, displayed schematically in Fig. \ref{fig:disp_sub_in_out}. We consider the modes of the field $\phi$. An identical analysis is valid for $\varphi$, up to the replacement of the $D \rightarrow E$.

\vspace{0.4cm}
\noindent{\bf $\bullet$ Mode \textbf{$u_{\omega,\phi}^{v,in}$}}
\vspace{0.4cm}

\noindent The ``in'' $v$-mode $u_{\omega,\phi}^{v,in}$ is defined by an initial unit-amplitude left-moving $v$-mode coming from the right ($\equiv u_{\omega,\phi}^{v,r}=D_v^re^{-i\omega t+ik_v^r x}$), which is partially transmitted into a $v$-mode in the left region ($u_{\omega,\phi}^{v,l}=D_v^le^{-i\omega t+ik_v^l x}$) with amplitude $A_{v}^l$ and partially reflected into a right-moving $u$-mode ($u_{\omega,\phi}^{u,r}=D_u^re^{-i\omega t+ik_u^r x}$) with amplitude $A_{u}^r$. The construction is not finished yet, as we need to include as well the decaying modes in the left and right regions ($u_{\omega,\phi}^{d,r(l)}=D_{\phi}^{r(l)}e^{-i\omega t+ik_{d(g)}^{r(l)} x}$) along with their amplitudes $A_d^r$ and $A_d^l$. In this way we have a total of four amplitudes which are uniquely determined by solving the following system of four equations
\begin{equation}
\label{modevin}
     \left( \begin{array}{c}
       A_v^l \\
       0 \\
       0\\
       A_d^l\\
     \end{array} \right)
   = M_{scatt}\left(
                \begin{array}{c}
                  1  \\
                  A_u^r \\
                  A_d^r\\
                  0
                \end{array}
              \right) \ .
\end{equation}
By treating $M_{scatt}$ perturbatively in the parameter $z_l\equiv \frac{\omega\xi_l}{c_l}$ we find, up to $O(z^2)$, the following solutions
\bea
A_v^l &=&\frac{2 \sqrt{c_l c_r}}{c_l+c_r} -\frac{i \sqrt{c_l}  \left(c_l-c_r\right){}^2 z _l}{ c_r^{3/2} \left(c_l+c_r\right)}+\frac{c_l \left(c_l-c_r\right){}^2 \left(c_l^2+c_r^2\right) z _l^2}{2  c_r^3 \left(c_l+c_r\right){}^2}\equiv T\ ,\\
A_u^r &=&\frac{c_l-c_r}{c_l+c_r}-\frac{i c_l  \left(c_l-c_r\right){}^2 z _l}{c_r^2 \left(c_l+c_r\right)}-\frac{c_l \left(c_l-c_r\right) \left(2 c_l^3-3 c_l^2 c_r+2 c_l c_r^2+c_r^3\right) z _l^2}{4  c_r^4 \left(c_l+c_r\right)}\equiv R\ , \\
A_d^l &=& \frac{\left(c_l-c_r\right) \sqrt{z _l}}{d_{\phi}^l \sqrt{c_r} \left(c_l+c_r\right)}-\frac{\left(c_l-c_r\right)\text{  }z _l^2}{2 d_{\phi}^l c_r^{5/2} \left(c_l+c_r\right)}\left[c_r{}^2+i\left(c_l{}^2+c_r{}^2-c_rc_l\right)\right]\equiv D_{L'}\ , \\
A_d^r &=& \frac{ c_l \left(-c_l+c_r\right) \sqrt{z _l}}{d_{\phi}^r c_r^{3/2} \left(c_l+c_r\right)}+\frac{ c_l^2 \left(c_l-c_r\right)\text{  }z _l^2}{2d_{\phi}^r c_r^{7/2} \left(c_l+c_r\right)}\left[c_l+i\left(c_l-2c_r\right)\right]\equiv D_{R'}\ .
\eea
In the limit $z_l \rightarrow 0$, we recover the results of \cite{nostro}. As we can see, the amplitudes of the asymptotic modes $A_v^l$ and $A_u^r$ develop an imaginary $O(z_l)$ contribution plus a real $O(z_l^2)$ one. These combine in such a way that the unitarity relation $|A_v^l|^2+|A_u^r|^2\equiv |R|^2+|T|^2=1$ is satisfied non trivially at $O(z_l^2)$, as
\bea
|A_v^l|^2 &=&\frac{4c_lc_r}{(c_l+c_r)^2}+\frac{\omega ^2 \left(c_l-c_r\right){}^2 \left(c_l^2+c_r^2\right) \xi _l^2}{2 c_l c_r^3 \left(c_l+c_r\right){}^2}\ ,\\
|A_u^r|^2 &=&\left(\frac{c_l-c_r}{c_l+c_r}\right)^2-\frac{\omega ^2 \left(c_l-c_r\right){}^2 \left(c_l^2+c_r^2\right) \xi _l^2}{2 c_l c_r^3 \left(c_l+c_r\right){}^2}\ .
\eea
Finally, note that, although the amplitudes of the decaying modes do not enter in the unitarity relation, they are part of the full mode and give contributions, for instance, in the computation of density-density correlations.

\vspace{0.4cm}
\noindent{\bf $\bullet$ Mode \textbf{$u_{\omega,\phi}^{u,in}$}}
\vspace{0.4cm}

\noindent The ``in'' $u$-mode $u_{\omega,\phi}^{u,in}$ is composed by an initial unit-amplitude right-moving $u$-mode $(u_{\omega,\phi}^{u,l}\equiv D_u^le^{-i\omega t+ik_u^lx})$ coming from the left, along with the transmitted $u$-mode ($u_{\omega}^{u,r}$) with amplitude $A_{u}^r$ and the reflected $v$-mode ($u_{\omega,\phi}^{v,l}$) with amplitude $A_{v}^l$. Here too we have decaying modes, with amplitudes $A_d^r,A_d^l$. All these amplitudes are obtained by solving
\begin{equation}
\label{modeuin} %multiply defined, used only in the appendix, I think this is the good one
     \left( \begin{array}{c}
       A_v^l \\
       1  \\
       0\\
       A_d^l\\
     \end{array} \right)
   = M_{scatt}\left(
                \begin{array}{c}
                  0  \\
                  A_u^r \\
                  A_d^r\\
                  0
                \end{array}
              \right) \
\end{equation}
and, up to $O(z_l^2)$, we have

\bea
  A_v^l &=& \frac{c_r-c_l}{c_l+c_r}-\frac{i   \left(c_l-c_r\right){}^2 z _l}{c_r \left(c_l+c_r\right)}+\frac{\left(c_l-c_r\right) \left(c_l^3+2 c_l^2 c_r-3 c_l c_r^2+2 c_r^3\right) z _l^2}{4  c_r^3 \left(c_l+c_r\right)}\equiv R'\ ,\\
A_u^r &=&\frac{2 \sqrt{c_l c_r}}{c_l+c_r}-i\frac{ \sqrt{c_l}  \left(c_l-c_r\right){}^2 z _l}{ c_r^{3/2} \left(c_l+c_r\right)}-\frac{\sqrt{c_l} \left(c_l-c_r\right){}^2 \left(c_l^2-4 c_l c_r+c_r^2\right) z _l^2}{8  c_r^{7/2} \left(c_l+c_r\right)}\equiv T'\ ,\\
A_d^l&=&\frac{  \left(c_l-c_r\right) \sqrt{z_l}}{d_{\phi}^l \sqrt{c_l} \left(c_l+c_r\right)}+\frac{  \left(c_l-c_r\right)\text{  }}{2d_{\phi}^l\sqrt{c_l}  c_r \left(c_l+c_r\right)}\left[-c_r+i\left(2c_l-c_r\right)\right]z_l^2\equiv D_L\ ,\\
  A_d^r&=&\frac{  \sqrt{c_l} \left(-c_l+c_r\right) \sqrt{z _l}}{d_{\phi}^r c_r \left(c_l+c_r\right)}+\frac{\sqrt{c_l}\left(c_l-c_r\right)}{2 d_{\phi}^r  c_r^3 \left(c_l+c_r\right)}\left[c_l{}^2+i\left(c_l{}^2+c_r{}^2-c_lc_r\right)\right]z _l^2\equiv D_R\ .
\eea
The unitarity condition for the asymptotic modes $|A_v^l|^2+|A_u^r|^2\equiv |R'|^2+|T'|^2=1$ is again non-trivially satisfied, as
\bea
|A_v^l|^2 &=& \left(\frac{c_r-c_l}{c_l+c_r}\right)^2-\frac{c_l \left(c_l-c_r\right){}^2 \left(c_l^2+c_r^2\right) z _l^2}{2c_r^3 \left(c_l+c_r\right){}^2}\ ,\\
|A_u^r|^2 &=&\frac{4c_lc_r}{(c_l+c_r)^2}+\frac{c_l\left(c_l-c_r\right){}^2 \left(c_l^2+c_r^2\right) z _l^2}{2c_r^3 \left(c_l+c_r\right){}^2} \ .
\eea

%The `out' modes can be constructed with a similar procedure, as we now briefly show (see Fig \ref{fig:disp_sub_in_out}).

\vspace{0.4cm}
\noindent{\bf $\bullet$ Mode \textbf{$u_{\omega,\phi}^{v,out}$}}
\vspace{0.4cm}

\noindent The ``out'' $v$-mode $u_{\omega,\phi}^{v,out}$ is made of a linear combination of initial right-moving ($u_{\omega,\phi}^{u,l}$) and left-moving ($u_{\omega,\phi}^{v,r}$) components, with amplitudes $A_u^l$ and $A_v^r$, producing a final left-moving $v$-component $(u_{\omega,\phi}^{v,l})$ of unit amplitude. The amplitudes, together with those of the associated decaying modes, are given by solving
\begin{equation}
\label{modevout}
     \left( \begin{array}{c}
       1 \\
       A_u^l  \\
       0\\
       A_d^l\\
     \end{array} \right)
   = M_{scatt}\left(
                \begin{array}{c}
                  A_v^r  \\
                  0 \\
                  A_d^r\\
                  0
                \end{array}
              \right) \
\end{equation}
and, at $O(z^2)$, one has
\bea
  A_u^l &=&\frac{c_r-c_l}{c_l+c_r}+\frac{i \left(c_l-c_r\right){}^2 z _l}{ c_r \left(c_l+c_r\right)}+\frac{ \left(c_l-c_r\right) \left(c_l^3+2 c_l^2 c_r-3 c_l c_r^2+2 c_r^3\right) z _l^2}{4  c_r^3 \left(c_l+c_r\right)}\equiv R^{'*}\ ,\\
A_v^r &=&\frac{2 \sqrt{c_l c_r}}{c_l+c_r} +\frac{i \sqrt{c_l}  \left(c_l-c_r\right){}^2 z _l}{ c_r^{3/2} \left(c_l+c_r\right)}-\frac{ \left(c_l-c_r\right){}^2 \left(c_l^2-4 c_l c_r+c_r^2\right) z _l^2}{8 c_r^{7/2} \left(c_l+c_r\right)}\equiv T^{'*}\ ,\\
A_d^l&=&\frac{ \left(c_l-c_r\right) \sqrt{z _l}}{d_{\phi}^l \sqrt{c_l}  \left(c_l+c_r\right)}-\frac{ \left(c_l-c_r\right)\text{  }z _l^2}{2d_{\phi}^l  c_r \left(c_l+c_r\right)}\left[c_r+i\left(2c_l-c_r\right)\right]\equiv D_{L}^*\ ,\\
  A_d^r&=&\frac{\sqrt{c_l} \left(-c_l+c_r\right) \sqrt{z _l}}{d_{\phi}^r c_r \left(c_l+c_r\right)}+\frac{\sqrt{c_l} \left(c_l-c_r\right)\text{  }z _l^2}{2 d_{\phi}^r  c_r^3 \left(c_l+c_r\right)}\left[c_l{}^2-i\left(c_l{}^2+c_r{}^2-c_rc_l\right)\right]\equiv D_R^*\ .
\eea
One can easily check that the unitarity relation is satisfied, as $|A_u^l|^2+|A_v^r|^2\equiv |R|^2+|T|^2=1$.

\vspace{0.4cm}
\noindent{\bf $\bullet$ Mode \textbf{$u_{\omega,\phi}^{u,out}$}}
\vspace{0.4cm}

\noindent We finally consider the mode $u_{\omega,\phi}^{u,out}$. This is defined by initial right-moving and left-moving components, with amplitudes $A_u^l$ and $A_v^r$, resulting now in a final right-moving $u$ component $(u_{\omega,\phi}^{u,r})$ of unit amplitude. The system of equations to be solved (taking into account the decaying modes) is
\begin{equation}
\label{modeuout}
     \left( \begin{array}{c}
       0 \\
       A_u^l  \\
       0\\
       A_d^l\\
     \end{array} \right)
   = M_{scatt}\left(
                \begin{array}{c}
                  A_v^r  \\
                  1 \\
                  A_d^r\\
                  0
                \end{array}
              \right) \ .
\end{equation}
and its solutions, up to $O(z^2)$, are

\bea
A_u^l &=&\frac{2 \sqrt{c_l c_r}}{c_l+c_r} +\frac{i \sqrt{c_l}  (c_l-c_r)^2 z_l}{c_r^{3/2} \left(c_l+c_r\right)}-\frac{\sqrt{c_l} \left(c_l-c_r\right){}^2 \left(c_l^2-4 c_l c_r+c_r^2\right) z_l^2}{8  c_r^{7/2} \left(c_l+c_r\right)}\equiv T^*\ ,\\
A_v^r &=&\frac{c_l-c_r}{c_l+c_r}+\frac{i c_l \left(c_l-c_r\right){}^2 z_l}{c_r^2 \left(c_l+c_r\right)}-\frac{c_l\left(c_l-c_r\right) \left(2 c_l^3-3 c_l^2 c_r+2 c_l c_r^2+c_r^3\right) z_l^2}{4 c_r^4 \left(c_l+c_r\right)} \equiv R^*\ ,\\
A_d^l&=&\frac{c_l \left(c_l-c_r\right) z_l}{d_{\phi}^l \sqrt{c_r} \left(c_l+c_r\right)}+\frac{\left(c_l-c_r\right)\text{  }z _l^2}{2 d_{\phi}^l  c_r^{5/2} \left(c_l+c_r\right)}\left[-c_r{}^2+i\left(c_l{}^2+c_r{}^2-c_lc_r\right)\right]\equiv D_{L^{'*}}\ ,\\
A_d^r&=&\frac{c_l \left(-c_l+c_r\right) z_l}{d_{\phi}^r c_r^{3/2} \left(c_l+c_r\right)}+\frac{  c_l^2 \left(c_l-c_r\right)\text{  }z_l^2}{2d_{\phi}^r c_r^{7/2} \left(c_l+c_r\right)}\left[c_l+i\left(2c_r-c_l\right)\right]\equiv D_{R^{'*}}\ ,
\eea
with unitarity condition $|A_u^l|^2+|A_v^r|^2=|R|^2+|T|^2=1$ satisfied up to $O(z_l^2)$.

Having constructed explicitly the complete ``in'' and ``out'' modes basis, we can now write the two alternative decompositions for the field operator $\hat\phi$
\begin{eqnarray}\non
\hat\phi&=&\int_{0}^{\infty}d\omega\Big[\hat a_{\omega}^{v,in(out)}u_{\omega,\phi}^{v,in(out)}(t,x)+\hat a_{\omega}^{u,in(out)}u_{\omega,\phi}^{u,in(out)}(t,x)+\\
&+&\hat a_{\omega}^{v,in(out)\dagger}u_{\omega,\varphi}^{v,in(out)*}(t,x)+\hat a_{\omega}^{u,in(out)\dagger}u_{\omega,\varphi}^{u,in(out)*}(t,x)\Big].
\end{eqnarray}
The relations between the  ``in'' and ``out'' modes are
\begin{eqnarray}\label{rt}
  u_{\omega,\phi}^{v,in} &=& Tu_{\omega,\phi}^{v,out}+Ru_{\omega,\phi}^{u,out}, \nonumber\\
  u_{\omega,\phi}^{u,in} &=& R'u_{\omega,\phi}^{v,out}+T'u_{\omega,\phi}^{u,out},
\end{eqnarray}
and are valid for all components of the modes basis, decaying modes included. This allows us to find
\begin{eqnarray}
  \hat a_{\omega}^{v,out} &=& T\hat a_{\omega}^{v,in}+R'\hat a_{\omega}^{u,in}, \nonumber\\
  \hat a_{\omega}^{u,out} &=& R\hat a_{\omega}^{v,in}+T'\hat a_{\omega}^{u,in}.
\end{eqnarray}

\vspace{0.4cm}
\noindent{\bf Density-density correlations}
\vspace{0.4cm}

\noindent The basic quantity that we want to study in detail later is the one-time, normalized, symmetric, two-point function of the density fluctuation
\begin{equation}\label{densitya}
G^{(2)}(t;x,x')\equiv \frac{1}{2n^2}\lim_{t\rightarrow t'}\langle {\rm in}|\{ \hat n^1 (t,x), \hat n^1 (t',x')\}|{\rm in}\rangle\ \ ,
\end{equation}
where $\{,\}$ denotes the anticommutator, and the the operator $\hat n^1\equiv n(\hat \phi + \hat \phi^{\dagger})$ (see eq. (\ref{dph})) can be expanded in the two equivalent ``in'' and ``out'' representations,
\bea
\hat n^1=n\int_0^{\infty}d\omega\left[\hat a_{\omega}^{v,in(out)}(u_{\omega,\phi}^{v,in(out)}+u_{\omega,\varphi}^{v,in(out)})+
\hat a_{\omega}^{u,in(out)}(u_{\omega,\phi}^{u,in(out)}+u_{\omega,\varphi}^{u,in(out)})+{\rm h.c.}\right]\ .
\eea
Thus, the general two-point function in (\ref{densitya}) explicitly reads
\bea\label{g2subesp}
&&\langle in|\{ \hat n^1 (t,x), \hat n^1 (t',x')\}| in\rangle|=n^2\int_0^{\infty}d\omega\left[(u_{\omega,\phi}^{v,in(out)}+u_{\omega,\varphi}^{v,in(out)})(t,x)(u_{\omega,\phi}^{v,in(out)*}+u_{\omega,\varphi}^{v,in(out)*})(t',x')\right.+\non\\
&&\left.+(u_{\omega,\phi}^{u,in(out)}+u_{\omega,\varphi}^{u,in(out)})(t,x)(u_{\omega,\phi}^{u,in(out)*}+u_{\omega,\varphi}^{u,in(out)*})(t',x')+{\rm c.c.}\right],
\eea
where
\begin{eqnarray}
u_{\omega,\phi}^{v,in}+u_{\omega,\varphi}^{v,in} &=& e^{-i\omega t}\left[(D_v^r+E_v^r)e^{ik_v^r(\omega)x}+R(D_u^r+E_u^r)e^{ik_u^r(\omega)x}+T(D_v^l+E_v^l)e^{ik_v^l(\omega)x}+\right.\nonumber\\
&+&\left.(D_{L'}^{\phi}d_{\phi}^l+D_{L'}^{\varphi}d_{\varphi}^l)e^{ik_g^l(\omega)x}+(D_{R'}^{\phi}d_{\phi}^r+D_{R'}^{\varphi}d_{\varphi}^r)e^{ik_d^r(\omega)x}\right],\nonumber\\
u_{\omega,\phi}^{u,in}+u_{\omega,\varphi}^{u,in} &=& e^{-i\omega t}\left[(D_u^l+E_u^l)e^{ik_u^l(\omega)x}+R'(D_v^l+E_v^l)e^{ik_v^l(\omega)x}+T'(D_u^r+E_u^r)e^{ik_u^r(\omega)x}+\right.\nonumber\\
&+&\left.(D_{L}^{\phi}d_{\phi}^l+D_{L}^{\varphi}d_{\varphi}^l)e^{ik_g^l(\omega)x}+(D_{R}^{\phi}d_{\phi}^r+D_{R}^{\varphi}d_{\varphi}^r)e^{ik_d^r(\omega)x}\right].
\end{eqnarray}

%
%\begin{eqnarray}
%\label{eq:relationinoutmo}
%  u_{\omega}^{v,in} &=& u_{\omega}^{v,r}+R u_{\omega}^{u,r}+T u_{\omega}^{v,l}\ , \nonumber\\
%  u_{\omega}^{u,in} &=& u_{\omega}^{u,l}+R' u_{\omega}^{v,l}+T' u_{\omega}^{u,r}\ ,
%  %  u_{\omega}^{u,out}&=& Tu_{\omega}^{u,in}+Ru_{\omega}^{v,in} \nonumber \\
%%  u_{\omega}^{v,out} &=& -Ru_{\omega}^{u,in}+Tu_{\omega}^{v,in}
%\end{eqnarray}

\noindent Let us consider, for instance, one point located in the left ($x<0$) region and one in the right ($x'>0$) one. Substituting the expressions above into (\ref{g2subesp}), we see that there are $u-u$ and $v-v$ contributions, while the $u-v$ term, being proportional to $R^*T+R'T'^*$, vanishes. Finally, the contribution coming from the decaying modes is subdominant. Therefore, the integral (\ref{g2subesp}) is well approximated by the hydrodynamic approximation, obtained for small $\omega$, namely
%
%\begin{eqnarray}
%&&\langle in| \{ \hat n^1 (t,x),\hat n^1 (t',x')\} |in\rangle= \int_0^{\infty}d\omega \Big[ T^{'*}u_w^{u,l}(t,x)u_w^{u,r *}(t',x') + \\
%&& T  u_w^{v,l}(t,x)u_w^{v,r *}(t',x') +  (R^*T+R'T^{'*})u_{\omega}^{v,l}(t,x)u_{\omega}^{u,r *}(t',x')  + c.c.\Big] = \nonumber \\
%&&-\frac{\hbar mc_rc_l}{2\pi n(c_r+c_l)} \ln [(t-t'-\frac{x}{v+c_l}+\frac{x'}{v+c_r})
%(t-t' -\frac{x}{v-c_l}+\frac{x'}{v-c_r})]  \ .\nonumber
%\end{eqnarray}

%We can now go to the hydrodynamical limit by expanding the expression above for small $\omega$:

\begin{eqnarray}
G^{(2)}(t;x,x')\simeq -\frac{\hbar}{2\pi mn(c_r+c_l)}\left[\frac{1}{(v-c_l)(v-c_r)\left(\frac{x}{c_l-v}+\frac{x'}{v-c_r}\right)^2}+\frac{1}{(v+c_l)(v+c_r)\left(-\frac{x}{v+c_l}+\frac{x'}{v+c_r}\right)^2}\right]\ .
\end{eqnarray}

\subsection{Subsonic-supersonic configuration}\label{supconf}

\begin{figure}[htbp]
\begin{center}
%\resizebox{!}{8cm}{\input{uhidro.eps}}
\resizebox{!}{5cm}{\includegraphics{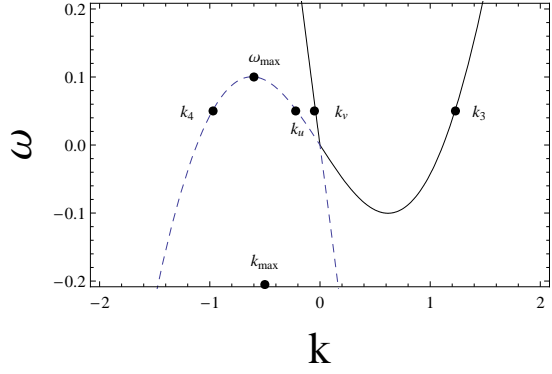}}
\caption{Dispersion relation in the supersonic case. Positive (negative) norm modes belong to the solid (dashed) line.}
\label{fig:sup_disp}
\end{center}
\end{figure}

\noindent Unlike the spatial step-like discontinuities studied in \cite{nostro} in the hydrodynamical limit,  dispersion effects allow us to study also configurations with supersonic regions. Since we are interested to model black hole-like systems, we shall consider the case where there are one subsonic and one supersonic region separated by a sharp jump in the speed of sound. Therefore we write $c(x)=c_l\theta(-x) + c_r \theta(x)$, where now $c_l<|v|$ and $c_r>|v|$. The modes in the subsonic region ($x>0$) are the same as in the previous subsection. In the supersonic ($x<0$) part the dispersion relation (\ref{nrela}) changes and it is represented in Fig.\  \ref{fig:sup_disp}.
We see that, for $\omega$ less than a certain value that we call $\omega_{max}$, there are now four real solutions, corresponding to four propagating modes. Two of them are present also in the hydrodynamical approximation, and, when expressed through the variable $z_l\equiv {\xi_l \omega\over c_l}$, they read (we omit the subscript $l$)
\begin{eqnarray}
&&k_v=\frac{\omega}{v-c}\left[1+\frac{c^3z^2}{8(v-c)^3}+O(z_l^2)\right]\ ,\nonumber\\
&&k_u=\frac{\omega}{v+c}\left[1-\frac{c^3z^2}{8(v+c)^3}+O(z_l^2)\right]\ ,
\end{eqnarray}
and, unlike in the subsonic case, they both move to the left, as $\frac{d\omega}{dk}|_{k_{u(v)}}<0$. The value $k_v$ belongs to the positive norm branch while $k_u$ to the negative norm one, as shown in Fig.\  \ref{fig:sup_disp}.
The other two values of $k$, called $k_3$ and $k_4$, exist because of dispersion, and are not perturbative in $\xi$. In fact
\bea\non
k_{3(4)}&=&\frac{\omega |v|}{c^2-v^2}\left[1-\frac{(c^2+v^2)c^4z^2}{4(c^2-v^2)^3}+O(z^4)\right]+(-)\frac{2\sqrt{v^2-c^2}}{c\xi}\left[1+\frac{(c^2+2v^2)c^4z^2}{8(c^2-v^2)^3}+O(z^4)\right]\ .
\eea
Comparing with the expressions (\ref{decaying}), we see that $k_3$ and $k_4$ are the analytic continuation for supersonic flows of the decaying and growing modes seen in the subsonic regime. These two modes (which belong respectively to the positive and negative norm branches of Fig \ref{fig:sup_disp}) both move to the right as $\frac{d\omega}{dk}|_{k_{3(4)}}>0$. This means that they are supersonic and able to propagate upstream, against the direction of the flow. The value of $\omega_{max}=\omega(k_{max})$ can be calculated explicitly by imposing $\frac{d\omega}{dk}|_{k=k_{max}}=0$, where
\bea
k_{max}=-\frac{1}{\xi}\left[\frac{v^2}{2c^2}-2+\frac{v}{2c}\sqrt{8+\frac{v^2}{c^2}}\right]^{1/2}\ .
\eea
One can easily check that $\omega_{max}$ and $k_{max}$ are well inside the non-perturbative region ($\sim 1/\xi$). When $\omega>\omega_{max}$, instead, we find again two real propagating modes ($k$ real) and two complex conjugate ones, corresponding to decaying and growing modes, just like in the subsonic case. Thus, for $\omega>\omega_{max}$ the analysis is the same as in the subsonic case, and so we omit it.

Let us now write the general solutions for $\phi$ and $\varphi$ in the left ($l$) and in the right ($r$) regions for $\omega<\omega_{max}$.  In the $l$-region we have
\begin{eqnarray}
  \phi_{\omega}^{l} &=& e^{-i\omega t}\left[D_v^{l}A_v^{l}e^{ik_v^{l}x}+D_u^{l}A_u^{l}e^{ik_u^{l}x}+D_{3}^{l}A_3^{l}e^{ik_3^{l}x}+D_{4}^{l}A_4^{l}e^{ik_4^{l}x}\right], \nonumber\\
  \varphi_{\omega}^{l} &=& e^{-i\omega t}\left[E_v^{l}A_v^{l}e^{ik_v^{l}x}+
  E_u^{l}A_u^{l}e^{ik_u^{l}x}+E_{3}^{l}A_3^{l}e^{ik_3^{l}x}+E_{4}^{l}A_4^{l}e^{ik_4^{l}x}\right] \nonumber\ ,
  \end{eqnarray}
while in the $r$-region we find
\begin{eqnarray}
  \phi_{\omega}^{r} &=& e^{-i\omega t}\left[D_v^{r}A_v^{r}e^{ik_v^{r}x}+D_u^{r}A_u^{r}e^{ik_u^{r}x}+d_{\phi}A_d^{r}e^{ik_d^{3}x}+G_{\phi}A_g^{r}e^{ik_g^{r}x}\right], \nonumber\\
  \varphi_{\omega}^{r} &=& e^{-i\omega t}\left[E_v^{r}A_v^{r}e^{ik_v^{r}x}+
  E_u^{r}A_u^{r}e^{ik_u^{r}x}+d_{\varphi}A_d^{r}e^{ik_d^{r}x}+G_{\varphi}A_g^{r}e^{ik_g^{r}x}\right]\ . \nonumber
  \end{eqnarray}
The $D$ and $E$ normalization coefficients of the propagating modes (four in the supersonic region and two in the subsonic region) are given by Eqs.\  (\ref{eq:normdispersion}). As before, the matching conditions (\ref{matchingaa}) can be written in the matrix form
\begin{equation}
%\label{eq:bogoc}
W_l\left(
     \begin{array}{c}
       A_v^l \\
       A_u^l \\
       A_3^l \\
       A_4^l \\
 \end{array}
   \right)=W_r\left(
                \begin{array}{c}
                  A_v^r \\
       A_u^r \\
       A_d^r \\
       A_g^r \\
                \end{array}
              \right),
\end{equation}
where $W_r$ is the same as Eq.\  (\ref{eq:wr}), while $W_l$ is given by
\begin{equation}
%\label{eq:wl} multiply defined!
W_l=\left(
     \begin{array}{cccc}
       D_v^l & D_u^l & D_{3}^l & D_{4}^l\\
       ik_v^lD_v^l & ik_u^lD_u^l & ik_3^lD_{3}^l & ik_4^lD_{4}^l \\
       E_v^l & E_u^l & E_{3}^l & E_{4}^l \\
       ik_v^lE_v^l & ik_u^lE_u^l & ik_3^lE_{3}^l & ik_4^lE_{4}^l  \\
\end{array}\right).
\end{equation}
Multiplying both sides by $W_l^{-1}$ we find
\begin{equation}
%\label{eq:matchingtc} no need for this label
     \left( \begin{array}{c}
       A_v^l \\
       A_u^l \\
       A_3^l \\
       A_4^l \\
     \end{array} \right)
   =M_{scatt} \left(
                \begin{array}{c}
                             A_v^r \\
       A_u^r \\
       A_d^r \\
       A_g^r \\
                \end{array}
              \right) \ ,
\end{equation}
where $M_{scatt}\equiv W_l^{-1} W_r$ encodes the scattering effects due to the matching conditions (\ref{matchingaa}). As in the previous subsection, we shall proceed to the construction of the ``in'' and ``out'' mode basis for this configuration. With these, we will construct the decompositions of the field $\hat \phi$ along with the the density-density correlations.

\vspace{0.4cm}
\noindent {\bf Construction of the  ``in'' and ``out'' basis}
\vspace{0.4cm}

\noindent We shall now construct the  ``in'' and ``out'' basis, which are now composed of three modes each, as shown in Fig.\ \ref{fig:sub_sup}. Below, we find the leading-order amplitudes of the various amplitudes. In Appendix C, we display the next-to-leading order terms for $u_{\omega,\phi}^{3,in}$ and $u_{\omega,\phi}^{4,in*}$ in order to show that  unitarity relations are non-trivially recovered.

\begin{figure}[htbp]
\begin{center}
\resizebox{!}{12cm}{\input{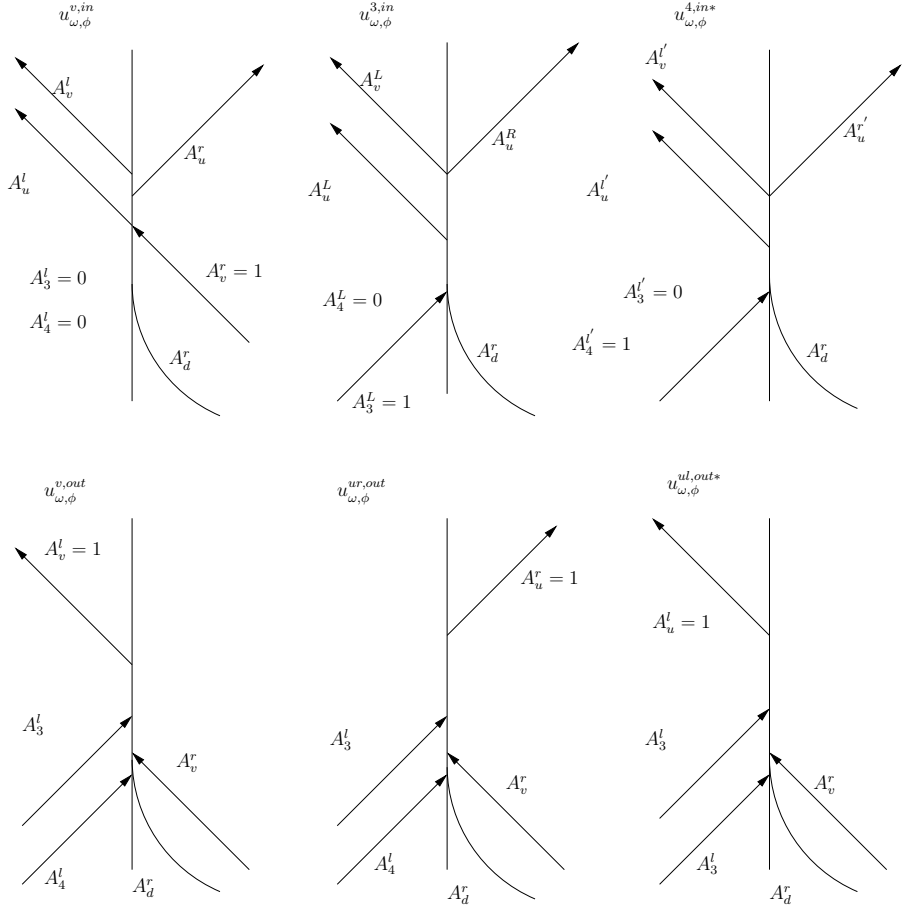}}
\caption{`in' and `out' basis in the subsonic-supersonic configuration.}
\label{fig:sub_sup}
\end{center}
\end{figure}

%\begin{figure}[htbp]
%\begin{center}
%%\resizebox{!}{8cm}{\input{uhidro.eps}}
%\resizebox{!}{8cm}{\includegraphics{supsub1in.eps}}
%\caption{The $u_{\omega}^{v,in}$ mode.}
%\label{fig:supsub1in}
%\end{center}
%\end{figure}
%
%\begin{figure}[htbp]
%\begin{center}
%%\resizebox{!}{8cm}{\input{uhidro.eps}}
%\resizebox{!}{8cm}{\includegraphics{supsub3in.eps}}
%\caption{The $u_{\omega}^{3,in}$ mode.}
%\label{fig:supsub3in}
%\end{center}
%\end{figure}
%
%\begin{figure}[htbp]
%\begin{center}
%%\resizebox{!}{8cm}{\input{uhidro.eps}}
%\resizebox{!}{8cm}{\includegraphics{supsub4in.eps}}
%\caption{The $u_{\omega}^{4,in}$ mode.}
%\label{fig:supsub4in}
%\end{center}
%\end{figure}

\newpage

\vspace{0.4cm}
\noindent {\bf $\bullet$ Mode $u_{\omega,\phi}^{v,in}$}
\vspace{0.4cm}

\noindent The mode $u_{\omega,\phi}^{v,in}$ is defined by an initial left-moving unit-amplitude component $(u_{\omega,\phi}^{v,r})$ coming from the subsonic region on the right, which generates a reflected right-moving mode ($u_{\omega,\phi}^{u,r}$) with amplitude $A_u^r$, together with the associated decaying mode with amplitude $A_d^r$. In addition, now there are two transmitted modes, one with positive norm ($u_{\omega,\phi}^{v,l}$) and the other with negative norm ($u_{\omega,\phi}^{u,l*}$), with amplitudes $A_{v}^l$ and $A_u^l$ respectively. These can be computed by solving the system of equations
\begin{equation}
%\label{modeuin} multiply defined
     \left( \begin{array}{c}
       A_v^l \\
       A_u^l \\
       0\\
       0\\
     \end{array} \right)
   = M_{scatt}\left(
                \begin{array}{c}
                  1  \\
                  A_u^r \\
                  A_d^r\\
                  0
                \end{array}
              \right) \ .
\end{equation}
The leading order $O(1)$ solution in a $z_l$ expansion is
\begin{eqnarray}\label{vin}\non
A_v^l &=& \sqrt{\frac{c_r}{c_l}}\frac{v-c_l}{v-c_r}\ , \\\non
A_u^r &=& \frac{v+c_r}{v-c_r}\ ,\\\non
A_u^l &=& \sqrt{\frac{c_r}{c_l}}\frac{v+c_l}{c_r-v}\ ,\\
A_d^r &=& \frac{c_l\sqrt{z_l}\sqrt{c_r(v^2-c_l^2)}}{\sqrt{2}d_{\phi}(v-c_l)(c_r^2-v^2)^{3/2}(c_r+c_l)}\left[\sqrt{c_r^2-v^2}\left(v+\sqrt{v^2-c_l^2}\right)+i\left(v\sqrt{v^2-c_l^2}+v^2-c_r^2\right)\right]\ .
\end{eqnarray}
The amplitudes of the propagating modes satisfy the unitarity condition $|A_v^l|^2+|A_u^r|^2-|A_u^l|^2=1$.

\vspace{0.4cm}
\noindent {\bf $\bullet$ Mode $u_{\omega,\phi}^{3,in}$}
\vspace{0.4cm}

\noindent The mode $u_{\omega,\phi}^{3,in}$ corresponds to a unit amplitude, supersonic positive norm right-moving plane wave from the left $(u_{\omega,\phi}^{3,l}\equiv D_3^le^{-i\omega t+ik_3^l(\omega)x})$, which is reflected into a positive norm ($u_{\omega,\phi}^{v,l}$) and a negative norm ($u_{\omega,\phi}^{u,l*}$) component with amplitudes $A_v^L$ and $A_u^L$ moving to the left. In addition, there is a transmitted right moving mode in the subsonic region ($u_{\omega,\phi}^{u,r}$) with amplitude $A_{u}^R$ and the decaying mode with amplitude $A_d^r$. By solving the system
\begin{equation}
%\label{modeuin} multiply defined
     \left( \begin{array}{c}
       A_v^L \\
       A_u^L\\
       1\\
       0\\
     \end{array} \right)
   = M_{scatt}\left(
                \begin{array}{c}
                  0  \\
                  A_u^R \\
                  A_d^R\\
                  0
                \end{array}
              \right) \ .
\end{equation}
we find, at leading order in $z_l$,
\begin{eqnarray}\label{3in}
  A_v^L &=& \frac{(v^2-c_l^2)^{3/4}(v+c_r)}{c_l^{3/2}\sqrt{2z_l}(c_l+c_r)\sqrt{c_r^2-v^2}}\left(\sqrt{c_r^2-v^2}+i\sqrt{v^2-c_l^2}\right),\nonumber \\
  A_u^R &=& \frac{\sqrt{2c_r}(v^2-c_l^2)^{3/4}(v+c_r)}{c_l\sqrt{z_l}(c_r^2-c_l^2)\sqrt{c_r^2-v^2}}\left(\sqrt{c_r^2-v^2}+i\sqrt{v^2-c_l^2}\right), \nonumber\\
  A_u^L &=& \frac{(v^2-c_l^2)^{3/4}(v+c_r)}{c_l^{3/2}\sqrt{2z_l}(c_l-c_r)\sqrt{c_r^2-v^2}}\left(\sqrt{c_r^2-v^2}+i\sqrt{v^2-c_l^2}\right),\nonumber\\
  A_d^R &=& \frac{(v^2-c_l^2)^{1/4}}{2d_{\phi}(v^2-c_r^2)}(v-i\sqrt{c_r^2-v^2}).
\end{eqnarray}
Note that the amplitudes of the propagating modes diverge in the $z_l\rightarrow 0$ limit, and that, at leading order in $z_l$, one has $|A_v^L|^2+|A_u^R|^2-|A_u^L|^2=0$. In order to check the unitarity condition $|A_v^l|^2+|A_u^r|^2-|A_u^l|^2=1$, we need the next-to-leading order expansion, which is displayed in Appendix C.

\newpage

\vspace{0.4cm}
\noindent {\bf $\bullet$ Mode $u_{\omega,\phi}^{4,in*}$}
\vspace{0.4cm}

\noindent The mode $u_{\omega,\phi}^{4,in*}$ (where $^*$  means that this is a negative norm mode) consists of an initial unit amplitude supersonic right-moving component from the left $(u_{\omega,\phi}^{4,l*}\equiv D_4^le^{-i\omega t+ik_4^l(\omega)x})$, generating a reflected positive left-moving norm mode ($u_{\omega,\phi}^{v,l}$) and negative norm left-moving mode ($u_{\omega,\phi}^{u,l}$) with amplitudes $A_v^{l'}$ and $A_u^{l'}$ respectively. Moreover, in the subsonic region one has a transmitted right-moving wave  ($u_{\omega,\phi}^{u,r}$) with amplitude $A_{u}^{r'}$, and a decaying mode with amplitude $A_d^{r'}$. By solving
\begin{equation}
%\label{modeuin}multiply defined
     \left( \begin{array}{c}
       A_v^{l'} \\
       A_u^{l'}\\
       0\\
       1\\
     \end{array} \right)
   = M_{scatt}\left(
                \begin{array}{c}
                  0  \\
                  A_u^{r'} \\
                  A_d^{r'}\\
                  0
                \end{array}
              \right) \
\end{equation}
we find, at leading order,
\bea\label{4in}\non
  A_v^{l'} &=& \frac{(v^2-c_l^2)^{3/4}(v+c_r)}{c_l^{3/2}\sqrt{2z_l}(c_l+c_r)\sqrt{c_r^2-v^2}}\left(\sqrt{c_r^2-v^2}-i\sqrt{v^2-c_l^2}\right)\ , \\\non
  A_u^{r'} &=& \frac{\sqrt{2c_r}(v^2-c_l^2)^{3/4}(v+c_r)}{c_l\sqrt{z_l}(c_r^2-c_l^2)\sqrt{c_r^2-v^2}}\left(\sqrt{c_r^2-v^2}-i\sqrt{v^2-c_l^2}\right) \ ,\\\non
  A_u^{l'} &=& \frac{(v^2-c_l^2)^{3/4}(v+c_r)}{c_l^{3/2}\sqrt{2z_l}(c_l-c_r)\sqrt{c_r^2-v^2}}\left(\sqrt{c_r^2-v^2}-i\sqrt{v^2-c_l^2}\right)\ ,\\
  A_d^{r'} &=& \frac{(v^2-c_l^2)^{1/4}(v^2-c_l^2+v\sqrt{v^2-c_l^2})}{2d_{\phi}(c_r^2-v^2)(c_l^2-v^2+v\sqrt{v^2-c_l^2})}(v-i\sqrt{c_r^2-v^2})\ .
\eea
As for $u_{\omega,\phi}^{3,in}$, the amplitudes of the propagating modes diverge when $z_l\rightarrow 0$ and at this level of approximation they satisfy $|A_v^l|^2+|A_u^r|^2-|A_u^l|^2=0$. The unitarity condition $|A_v^l|^2+|A_u^r|^2-|A_u^l|^2=-1$ is checked in the Appendix C by considering the next to leading order terms.

The construction of the ``out'' modes proceeds similarly. These are $u_{\omega,\phi}^{v,out}$, $u_{\omega,\phi}^{ur,out}$ (of positive norm) and $u_{\omega,\phi}^{ul,out*}$ (of negative norm), which are composed by appropriate combinations of initial right-moving and left-moving components (plus the associated decaying mode). These generate respectively, unit amplitudes $u_{\omega,\phi}^{v,l}$, $u_{\omega,\phi}^{u,r}$, and $u_{\omega,\phi}^{u,l*}$. More in detail, we have the following cases.

%\begin{figure}[htbp]
%\begin{center}
%%\resizebox{!}{8cm}{\input{uhidro.eps}}
%\resizebox{!}{8cm}{\includegraphics{supsub1out.eps}}
%\caption{Supersonic-subsonic case: The $u_{\omega}^{v,out}$ mode.}
%\label{fig:supsub1out}
%\end{center}
%\end{figure}
%
%\begin{figure}[htbp]
%\begin{center}
%%\resizebox{!}{8cm}{\input{uhidro.eps}}
%\resizebox{!}{8cm}{\includegraphics{supsub2out.eps}}
%\caption{Supersonic-subsonic case: The $u_{-\omega}^{ul,out*}$ mode.}
%\label{fig:supsub2out}
%\end{center}
%\end{figure}
%
%\begin{figure}[htbp]
%\begin{center}
%%\resizebox{!}{8cm}{\input{uhidro.eps}}
%\resizebox{!}{8cm}{\includegraphics{supsub3out.eps}}
%\caption{Supersonic-subsonic case: The $u_{\omega}^{ur,out}$ mode.}
%\label{fig:supsub3out}
%\end{center}
%\end{figure}

%
%\bigskip
%{\center{\bf Mode \textbf{$u_{\omega,`\phi}^{v,out}$}}}
%\bigskip\\

\vspace{0.4cm}
\noindent {\bf $\bullet$ Mode $u_{\omega,\phi}^{v,out}$}
\vspace{0.4cm}

\noindent In this case, one needs to solve the system
\begin{equation}
%\label{modeuin} multiply defined
     \left( \begin{array}{c}
       1 \\
       0  \\
       A_3^l\\
       A_4^l\\
     \end{array} \right)
   = M_{scatt}\left(
                \begin{array}{c}
                  A_v^r  \\
                  0 \\
                  A_d^r\\
                  0
                \end{array}
              \right) \ ,
\end{equation}
which yields, at leading order,
\begin{eqnarray}\label{vout}\non
  A_3^l&=&\frac{(v^2-c_l^2)^{3/4}\sqrt{c_r^2-v^2}}{\sqrt{2z_l}c_l^{3/2}(c_r-v)(c_r+c_l)}(\sqrt{c_r^2-v^2}-i\sqrt{v^2-c_l^2})=A_v^{l'}\ ,   \\\non
  A_4^l&=&\frac{(v^2-c_l^2)^{3/4}\sqrt{c_r^2-v^2}}{\sqrt{2z_l}c_l^{3/2}(c_r-v)(c_r+c_l)}(\sqrt{v^2-c_l^2}-i\sqrt{c_r^2-v^2})=iA_v^L\ , \\\non
  A_v^r&=&\sqrt{\frac{c_r}{c_l}}\frac{c_l-v}{c_r-v}=A_v^l\ ,\\
  A_d^r&=&\frac{(v^2-c_l^2)(c_r^2-v^2+v\sqrt{v^2-c_r^2})}{\sqrt{2}d_{\phi}(c_r-v)(c_l+c_r)\sqrt{c_l(v^2-c_r^2)}c_l\sqrt{z_l}}\ .
\end{eqnarray}

\vspace{0.4cm}
\noindent {\bf $\bullet$ Mode $u_{\omega,\phi}^{ur,out}$}
\vspace{0.4cm}

\noindent In this case the system to solve is
\begin{equation}
%\label{modeuin} multiply defined
     \left( \begin{array}{c}
       0 \\
       A_u^l  \\
       0\\
       A_4^l\\
     \end{array} \right)
   = M_{scatt}\left(
                \begin{array}{c}
                  A_v^r  \\
                  1 \\
                  A_d^r\\
                  0
                \end{array}
              \right) \ ,
\end{equation}
and the solutions are
\begin{eqnarray}\label{urout}
  A_3^l&=&\frac{\sqrt{2c_r}(v^2-c_l^2)^{3/4}(v+c_r)}{\sqrt{z_l}c_l(c_r^2-c_l^2)\sqrt{c_r^2-v^2}}(\sqrt{c_r^2-v^2}-i\sqrt{v^2-c_l^2})=A_u^{r'}\ , \nonumber \\
  A_4^l&=&\frac{\sqrt{2c_r}(v^2-c_l^2)^{3/4}(v+c_r)}{\sqrt{z_l}c_l(c_r^2-c_l^2)\sqrt{c_r^2-v^2}}(-\sqrt{c_r^2-v^2}-i\sqrt{v^2-c_l^2})=-A_u^R\ ,\nonumber \\
  A_v^r&=&\frac{v+c_r}{v-c_r}=A_u^r\ ,\nonumber \\
  A_d^r&=&-i\frac{\sqrt{2}c_r(v^2-c_l^2)(c_r^2-v^2+v\sqrt{v^2-c_r^2})}{d_{\phi}(c_r-v)^{3/2}(c_r^2-c_l^2)\sqrt{c_r(v+c_r)}c_l\sqrt{z_l}}\ .
\end{eqnarray}

\vspace{0.4cm}
\noindent {\bf $\bullet$ Mode $u_{\omega,\phi}^{ul,out*}$}
\vspace{0.4cm}

\noindent In this final case, the system is
\begin{equation}
%\label{modeuin} multiply defined
     \left( \begin{array}{c}
       0\\
       1\\
       A_3^l\\
       A_4^l\\
     \end{array} \right)
   = M_{scatt}\left(
                \begin{array}{c}
                  A_v^r  \\
                  0 \\
                  A_d^r\\
                  0
                \end{array}
              \right), \
\end{equation}
and the solutions read
\begin{eqnarray}\label{ulout}\non
  A_3^l&=&\frac{(v^2-c_l^2)^{3/4}\sqrt{c_r^2-v^2}}{\sqrt{2z_l}c_l^{3/2}(v-c_r)(c_r-c_l)}(\sqrt{c_r^2-v^2}-i\sqrt{v^2-c_l^2})=-A_u^{l'}\ , \\\non
  A_4^l&=&\frac{(v^2-c_l^2)^{3/4}\sqrt{c_r^2-v^2}}{\sqrt{2z_l}c_l^{3/2}(v-c_r)(c_r-c_l)}(-\sqrt{c_r^2-v^2}-i\sqrt{v^2-c_l^2})=A_u^L\ , \\\non
  A_v^r&=&\sqrt{\frac{c_r}{c_l}}\frac{c_l+v}{v-c_r}=-A_u^l\ ,\\
  A_d^r&=&\frac{(c_l^2-v^2)(v^2-c_r^2-v\sqrt{v^2-c_r^2})}{\sqrt{2}d_{\phi}(v-c_r)(c_l-c_r)\sqrt{c_l(v^2-c_r^2)}c_l\sqrt{z_l}}\ .
\end{eqnarray}
With these results, we are able to write down the relations between the ``in'' and ``out'' modes
\begin{eqnarray}\label{eq:outinaa}\non
  u_{\omega,\phi}^{v,in}&=& A_v^l u_{\omega,\phi}^{v,out}+A_u^r u_{\omega,\phi}^{ur,out}+A_u^l u_{\omega,\phi}^{ul,out*}\ , \\\non
  u_{\omega,\phi}^{3,in} &=& A_v^L u_{\omega,\phi}^{v,out}+A_u^R u_{\omega,\phi}^{ur,out}+A_u^L u_{\omega,\phi}^{ul,out*}\ , \\
  u_{\omega,\phi}^{4,in*} &=& A_v^{l'} u_{\omega,\phi}^{v,out}+A_u^{r'} u_{\omega,\phi}^{ur,out}+A_u^{l'} u_{\omega,\phi}^{ul,out*}\ ,
\end{eqnarray}
We note that, unlike the subsonic case (\ref{rt}), we now have combinations of both positive and negative norm modes. Because of this, the two decompositions (we restrict our analysis to the case $\omega<\omega_{max}$ because it is the relevant one for our subsequent discussion) are given by
\begin{eqnarray}
\hat\phi&=&\int_{0}^{\omega_{max}}d\omega\left[\hat a_{\omega}^{v,in}u_{\omega,\phi}^{v,in}+\hat a_{\omega}^{3,in}u_{\omega,\phi}^{3,in}+\hat a_{\omega}^{4,in}u_{\omega,\phi}^{4,in}+\hat a_{\omega}^{v,in\dagger}u_{\omega,\varphi}^{v,in*}+\hat a_{\omega}^{3,in\dagger}u_{\omega,\varphi}^{3,in*}+\hat a_{\omega}^{4,in\dagger}u_{\omega,\varphi}^{4,in*}\right]\ ,\\\non
\hat\phi&=&\int_{0}^{\omega_{max}}d\omega\left[\hat a_{\omega}^{v,out}u_{\omega,\phi}^{v,out}+\hat a_{\omega}^{ur,out}u_{\omega,\phi}^{ur,out}+\hat a_{\omega}^{ul,out}u_{\omega,\phi}^{ul,out}+\hat a_{\omega}^{v,out\dagger}u_{\omega,\varphi}^{v,out*}+\hat a_{\omega}^{ur,out\dagger}u_{\omega,\varphi}^{ur,out*}+\hat a_{\omega}^{ul,out\dagger}u_{\omega,\varphi}^{ul,out*}\right]\ ,\\
\end{eqnarray}
and they are inequivalent. This can be easily seen by using (\ref{eq:outinaa}) to find the relation between the two families of $\hat a$ and $\hat a^{\dagger}$ operators
\begin{eqnarray}\label{eq:outinoperatorsa}
  \hat a_{\omega}^{v,out} &=& A_v^l \hat a_{\omega}^{v,in}+A_v^L \hat a_{\omega}^{3in}+A_v^{l'} \hat a_{\omega}^{4in\dagger},\nonumber \\
  \hat a_{\omega}^{ur,out} &=& A_u^r \hat a_{\omega}^{v,in}+A_u^R \hat a_{\omega}^{3in}+A_u^{r'} \hat a_{\omega}^{4in\dagger},\nonumber \\
  \hat a_{\omega}^{ul,out\dagger} &=& A_u^l \hat a_{\omega}^{v,in}+A_u^L \hat a_{\omega}^{3in}+A_u^{l'}\hat  a_{\omega}^{4in\dagger}.
\end{eqnarray}
The fact that the r.h.s. of these relations contain both creation and annihilation operators makes it clear that the two decompositions do not share the same vacuum state ($|\rm in\rangle\neq|\rm out\rangle$).

\vspace{0.4cm}
\noindent{\bf Density-density correlations}
\vspace{0.4cm}

\noindent To compute the normalized density-density correlation analogous to Eq.\  (\ref{densitya}), we first expand the operator $\hat n^1$  in the ``out'' decomposition
%We compute the density-density correlation
%\begin{equation}\label{density}
%\langle in| n_1 (t,x) n_1 (t',x')|in\rangle,
%\end{equation}
%where \begin{equation}\label{n1}n_1(t,x)=\hat{\phi}(t,x)+\hat{\phi}^{+}(t,x).\end{equation}
%In this case it is covenient to expand $\hat{\phi}$ in the relevant ``out'' basis
%\begin{equation}
%\hat{\phi}=\int_0^{\omega_{max}}d\omega\left[\hat{\phi_{\omega}}+\hat{\varphi_{\omega}^{\dagger}}\right],
%\end{equation}
%where
%\begin{eqnarray}
%% \nonumber to remove numbering (before each equation)
%  \hat{\phi_{\omega}} &=& a_{\omega}^{v,out}u_{\omega,\phi}^{v,out}+a_{\omega}^{ur,out}u_{\omega,\phi}^{ur,out}+a_{\omega}^{ul,out+}u_{\omega,\varphi}^{ul,out*}, \\
%   \hat{\varphi_{\omega}}^{\dagger} &=& a_{\omega}^{v,out+}u_{\omega,\varphi}^{v,out*}+a_{\omega}^{ur,out+}u_{\omega,\varphi}^{ur,out*}+a_{\omega}^{ul,out}u_{\omega,\phi}^{ul,out}.
%\end{eqnarray}
%
\begin{eqnarray}
\hat n^1(t,x)=n\int_{0}^{\omega_{max}}\left[\hat a_{\omega}^{v,out}(u_{\omega,\phi}^{v,out}+u_{\omega,\varphi}^{v,out})+
\hat a_{\omega}^{ur,out}(u_{\omega,\phi}^{ur,out}+u_{\omega,\varphi}^{ur,out})+\hat a_{\omega}^{ul,out}(u_{\omega,\phi}^{ul,out}+u_{\omega,\varphi}^{ul,out})+h.c.\right]\ ,
\end{eqnarray}
and we use the relation between the ``in'' and ``out'' operators (\ref{eq:outinoperatorsa}).
This gives the following two-point function in the $| in\rangle$ state
\begin{eqnarray}\non
&&\langle  in|\{ \hat n^1 (t,x), \hat n^1 (t',x')\} | in\rangle=\\\non
&&n^2\int_0^{\omega_{max}}d\omega\left\{\left[A_v^l(u_{\omega,\phi}^{v,out}+u_{\omega,\varphi}^{v,out})+A_u^r(u_{\omega,\phi}^{ur,out}+u_{\omega,\varphi}^{ur,out})+A_u^l(u_{\omega,\phi}^{ul,out*}+u_{\omega,\varphi}^{ul,out*})\right](t,x)\right.\times\\
&&\left.\times\left[A_v^{l*}(u_{\omega,\phi}^{v,out*}+u_{\omega,\varphi}^{v,out*})+A_u^{r*}(u_{\omega,\phi}^{ur,out*}+u_{\omega,\varphi}^{ur,out*})+A_u^{l*}(u_{\omega,\phi}^{ul,out}+u_{\omega,\varphi}^{ul,out})\right](t',x')+\right.\nonumber\\
&&\left.+\left[A_v^L(u_{\omega,\phi}^{v,out}+u_{\omega,\varphi}^{v,out})+A_u^R(u_{\omega,\phi}^{ur,out}+u_{\omega,\varphi}^{ur,out})+A_u^L(u_{\omega,\phi}^{ul,out*}+u_{\omega,\varphi}^{ul,out*})\right](t,x)\right.\times\nonumber\\
&&\left.\times\left[A_v^{L*}(u_{\omega,\phi}^{v,out*}+u_{\omega,\varphi}^{v,out*})+A_u^{R*}(u_{\omega,\phi}^{ur,out*}+u_{\omega,\varphi}^{ur,out*})+A_u^{L*}(u_{\omega,\phi}^{ul,out}+u_{\omega,\phi}^{ul,out})\right](t',x')+\right.\nonumber\\
&&\left.+\left[A_v^{l'*}(u_{\omega,\phi}^{v,out*}+u_{\omega,\varphi}^{v,out*})+A_u^{r'*}(u_{\omega,\phi}^{ur,out*}+u_{\omega,\varphi}^{ur,out*})+A_u^{l'*}(u_{\omega,\phi}^{ul,out}+u_{\omega,\varphi}^{ul,out})\right](t,x)\right.\times\nonumber\\
&&\left.\times\left[A_v^{l'}(u_{\omega,\phi}^{v,out}+u_{\omega,\varphi}^{v,out})+A_u^{r'}(u_{\omega,\phi}^{ur,out}+u_{\omega,\varphi}^{ur,out})+A_u^{l'}(u_{\omega,\phi}^{ul,out*}+u_{\omega,\varphi}^{ul,out*})\right](t',x')+c.c.\right\},
\end{eqnarray}
where, explicitly,
\bea
  u_{\omega,\phi}^{v,out}+u_{\omega,\varphi}^{v,out}&=&e^{-i\omega t}\Big[ (D_v^l+E_v^l)e^{ik_v^l(\omega)x}+A_v^r(D_v^r+E_v^r)e^{ik_v^r(\omega)x}+\\\non
  &+&A_3^l(D_3^l+E_3^l)e^{ik_3^l(\omega)x}+A_4^l(D_4^l+E_4^l)e^{ik_4^l(\omega)x}+A_d^r(d_{\phi}^r+d_{\varphi}^r)e^{ik_d^r(\omega)x}\Big]\ ,\\
  u_{\omega,\phi}^{ur,out}+u_{\omega,\varphi}^{ur,out}&=&e^{-i\omega t}\Big[ (D_u^r+E_u^r)e^{ik_u^r(\omega)x}+A_v^r(D_v^r+E_v^r)e^{ik_v^r(\omega)x}+\\
  &+&A_3^l(D_3^l+E_3^l)e^{ik_3^l(\omega)x}+A_4^l(D_4^l+E_4^l)e^{ik_4^l(\omega)x}+A_d^r(d_{\phi}^r+d_{\varphi}^r)e^{ik_d^r(\omega)x}\Big]\ ,\non\\
  u_{\omega,\phi}^{ul,out*}+u_{\omega,\varphi}^{ul,out*}&=&e^{-i\omega t}\Big[ (D_u^l+E_u^l)e^{ik_u^l(\omega)x}+A_v^r(D_v^r+E_v^r)e^{ik_v^r(\omega)x}+\\
  &+&A_3^l(D_3^l+E_3^l)e^{ik_3^l(\omega)x}+A_4^l(D_4^l+E_4^l)e^{ik_4^l(\omega)x}+A_d^r(d_{\phi}^r+d_{\varphi}^r)e^{ik_d^r(\omega)x}\Big]\ .\non
\eea
The coefficients $A_v^l, A_v^r, A_3^l, A_4^l$ and $A_d^r$ are given, respectively, in (\ref{vout}), (\ref{urout}) and (\ref{ulout}).
%From now on we shall consider only stationary correlations.
The analysis of the main correlation signals has already been performed in \cite{recatipavloff}. We are interested in the correlation between $u_{\omega}^{u,r}$ and $u_{\omega}^{u,l*}$ because this represents the main signal due to the Hawking effect (correlation between the Hawking quanta and their partners). We take $x$ ($x'$) in the left (right) region and evaluate the following integral
\begin{eqnarray}\non
&&\langle in|\{ \hat n^1 (t,x), \hat n^1 (t',x')\}|in\rangle|(u_{\omega}^{u,r}\leftrightarrow u_{\omega}^{u,l*})=n^2\int_0^{\omega_{max}}d\omega\left[A_u^{l'*}A_u^{r'}(u_{\omega,\phi}^{u,l}+u_{\omega,\varphi}^{u,l})(t,x)(u_{\omega,\phi}^{u,r}+u_{\omega,\varphi}^{u,r})(t',x')+\right.\\
&&\left.+(A_u^{l}A_u^{r*}+A_u^{L}A_u^{R*})(u_{\omega,\phi}^{u,l*}+u_{\omega,\varphi}^{u,l*})(t,x)(u_{\omega,\phi}^{u,r*}+u_{\omega,\varphi}^{u,r*})(t',x')+{\rm c.c.}\right].
%\int_0^{\omega_{max}}d\omega\left[A_2^{l'*}A_2^{r'}(D+E)|_{k_2^l}(D+E)|_{k_2^r}u_{-\omega}^{2lout}(t,x)u_{\omega}^{2rout}(t,x')\right]
\end{eqnarray}
The values of the above amplitudes are given in (\ref{vin}), (\ref{3in}) and (\ref{4in}). We also take into account that
\begin{equation}
[a_{\omega}^{ur,out},a_{\omega}^{ul,out}]=0\Rightarrow A_u^{l*}A_u^r+A_u^{L*}A_u^R-A_u^{l'*}A_u^{r'}=0,
\end{equation}
where we have used the relation between the ``in'' and ``out'' operators given in (\ref{eq:outinoperatorsa}). The term $A_u^{l*}A_u^r$ is subleading with respect to the other two terms, which go as $O(1/\omega)$, given that the main contribution to the integral above is valid for small $\omega$. Note also that the products $A_u^{L*}A_u^R$ (and $A_u^{l'*}A_u^{r'}$) are real at leading order. Therefore we have
\begin{eqnarray}
\langle {\rm in}|\{ \hat n^1 (t,x), \hat n^1 (t',x')\}|{\rm in}\rangle|(u_{\omega}^{u,r}\leftrightarrow u_{\omega}^{u,l*})\sim 4n^2\int_0^{\omega_{max}}\!\!\!d\omega\left\{A_u^{l'*}A_u^{r'}\,\,{\rm Re}\left[(u_{\omega,\phi}^{u,l}+u_{\omega,\varphi}^{u,l})(t,x)(u_{\omega,\phi}^{u,r}+u_{\omega,\varphi}^{u,r})(t',x')\right]\right\},
\end{eqnarray}
and, at equal times, the normalized two-point function is
\begin{eqnarray}\label{correlation_spatial}
G^{(2)}(t;x,x')(u_{\omega}^{u,r}\leftrightarrow u_{\omega}^{u,l*})\sim-\frac{1}{4\pi n}\frac{(v^2-c_l^2)^{3/2}}{c_l(v+c_l)(v-c_r)(c_r-c_l)}\frac{\sin\left[\omega_{max}(\frac{x'}{v+c_r}-\frac{x}{v+c_l})\right]}{\frac{x'}{v+c_r}-\frac{x}{v+c_l}}\ .
\end{eqnarray}
This result, which coincides with the one given in \cite{recatipavloff}, gives an estimate of the Hawking signal in correlations only for stationary configurations. Our aim is to perform a similar construction, but for acoustic black hole-like configurations which are formed at some time $t_0$, along the lines of  the numerical analysis presented in \cite{fnum}.

%%%%%%%%%%%%%%%%%%%%%%%%%%%%%%%%%%%%%%%%%%%%%%

\section{Step-like discontinuities in $t$ (homogenous case)}\label{temporal}

\noindent In this section, we study correlation functions in the case of temporally formed step-like discontinuities between homogeneous condensates. In subsection \ref{temporal_sub} we consider condensates which remain subsonic at all times. In subsection \ref{temporal_sup}  we turn to the more relevant case when the final condensate is supersonic.

\subsection{Subsonic configurations}\label{temporal_sub}

\noindent We consider a step-like discontinuity in $t$ (say, at $t=0$), separating two infinite homogeneous condensates: $c(t)=c_{in}\theta(-t)+c_{out}\theta(t)$. In this section we consider $|v|<c_{in(out)}$ so that the condensate is subsonic at all times. The aim is to determine the mode propagation at all times, and to define the ``in'' and ``out'' mode basis. The appropriate decompositions of our field $\hat \phi$ will be given afterwards.

The general solutions in the ``in'' ($t<0$) and ``out'' ($t>0$) regions describing the fields $\phi$ and $\varphi$ are of the form $De^{-iwt+ikx}$ and $Ee^{-iwt+ikx}$. The boundary conditions at $t=0$ require us to work at fixed $k$. Therefore we write
\bea
\phi_{k}=D(k)e^{-iw(k)t+ikx}\ ,\qquad\varphi_{k}=E(k)e^{-iw(k)t+ikx}\ ,
\eea
for which Eqs. (\ref{cde}) become
\begin{eqnarray}\label{eq:primerorden22aa}
\left[-(\omega-vk)+\frac{c\xi k^2}{2}+\frac{c}{\xi}\right]D(k)&=&-\frac{c}{\xi}E(k)\ , \nonumber\\
\left[(\omega-vk)+\frac{c\xi k^2}{2}+\frac{c}{\xi}\right]E(k)&=&-\frac{c}{\xi}D(k)\ ,
\end{eqnarray}
while the normalization condition (\ref{nor}) yields
\begin{equation}\label{nodi}
 |D(k)|^2 - |E(k)|^2=\frac{1}{2\pi \hbar n}\ .
 \end{equation}
The combination of Eqs. (\ref{eq:primerorden22aa}) gives rise to the non-linear dispersion relation (\ref{nrela}) represented in Fig 1, and to the normalization coefficients
\begin{eqnarray}\label{eq:normdispersionk}
   &&D(k) =  \frac{\omega -v k+\frac{c\xi k^2}{2}}{\sqrt{4\pi \hbar n c\xi k^2\left| (\omega-vk) \right| }}\ ,\nonumber\\
  &&E(k) = -\frac{\omega -v k-\frac{c\xi k^2}{2}}{\sqrt{4\pi \hbar nc\xi k^2\left| (\omega-vk)  \right| }}\ .
\end{eqnarray}
Here, $\omega=\omega(k)$ corresponds to the two real solutions to Eq.\  (\ref{nrela}), which is quadratic in $\omega$ at fixed $k$.  These read
\begin{eqnarray}
\omega_+(k)&=&vk+\sqrt{c^2k^2+\frac{c^2k^4\xi^2}{4}}\ ,\nonumber\\
\omega_-(k)&=&vk-\sqrt{c^2k^2+\frac{c^2k^4\xi^2}{4}}\ ,
\end{eqnarray}
where $\omega_+(k)$ corresponds to the positive norm branch, and $\omega_-(k)$ to the negative norm one. Note that there are no normalizable mode solutions with complex $k$, because in the infinite homogeneous ``in'' and ``out'' regions they would correspond to modes which decay on one side but grow without bound on the other.  Therefore, at fixed $k$, the general decompositions of $\phi$ and $\varphi$ in the ``out'' and ``in'' regions are
\bea
  \phi_{k}^{out(in)}&=&e^{ik x}\left[D_{out(in)}^+(k)A_{out(in)}e^{-i\omega_+^{out(in)}(k)t}+D_{out(in)}^-(k)B_{out(in)}e^{-i\omega_-^{out(in)}(k)t}\right], \\
  \varphi_{k}^{out(in)}&=&e^{ik x}\left[E_{out(in)}^+(k)A_{out(in)}e^{-i\omega_+^{out(in)}(k)t}+E_{out(in)}^-(k)B_{out(in)}e^{-i\omega_-^{out(in)}(k)t}\right].
\eea
For $k>0\,\,(<0)$ we have a positive norm right-moving (left-moving) mode ($\omega=\omega_+(k)$) and a negative norm left-moving (right-moving) one ($\omega=\omega_-(k)$).
According to (\ref{cde}), the matching conditions at $t=0$ are
\begin{equation}\label{matching}
[\phi]=0, \,[\varphi]=0\ ,
\end{equation}
which can be written in matrix form
\begin{equation}
%\label{eq:bogoc}
W_{out}\left(
     \begin{array}{c}
       A_{out} \\
       B_{out} \\
     \end{array}
   \right)=W_{in}\left(
                \begin{array}{c}
                  A_{in} \\
                  B_{in}\\
                \end{array}
              \right),
\end{equation}
where
\begin{equation}
W_{out(in)}=\left(
     \begin{array}{cc}
       D_{out(in)}^+(k) & D_{out(in)}^-(k)\\
       E_{out(in)}^+(k) & E_{out(in)}^-(k) \\
     \end{array}
   \right).
\end{equation}
Multiplying both sides by $W_{out}^{-1}$ we find
 \begin{equation}
\label{eq:matchingtc}
     \left( \begin{array}{c}
       A_{out} \\
       B_{out} \\
     \end{array} \right)
   =M_{bog} \left(
                \begin{array}{c}
                  A_{in} \\
                  B_{in} \\
                \end{array}
              \right) \ .
\end{equation}
Explicitly, the Bogoliubov matrix $M_{bog}\equiv W_{out}^{-1} W_{in}$ reads
 \begin{equation}
\label{eq:bogoliubovda}
M_{bog}=\frac{1}{2\sqrt{\Omega_{in}\Omega_{out}}}\left(
            \begin{array}{cc}
               \Omega_{in}+\Omega_{out} &
               \Omega_{in}-\Omega_{out} \\
               \Omega_{in}-\Omega_{out} &
               \Omega_{in}+\Omega_{out} \\
            \end{array}
          \right) \ ,
\end{equation}
where we define $\Omega^{out(in)}=|\omega^{out(in)}-vk|$. For $v=0$ we recover the formulas given in \cite{ftemp}.

%We shall consider two different configurations for the temporal step-like discontinuity: where the `in (`out') regions are subsonic (subsonic) and subsonic (supersonic).

%
%\bigskip
%{\subsection{\bf The subsonic case}}\label{temporal_sub}
%\bigskip
%
%
%In this case both the `in' and `out' regions are subsonic ($c_{in}>|v|$ and $c_{out}>|v|$).
%The two real roots in both regions are given by a left-mover ($k_v^{in(out)}$) and a right-mover ($k_u^{in(out)}$), whose perturbative expressions are given in (\ref{uv}).
%
%\bigskip
%{\center{\bf Connecting the `in' and `out' basis}}
%\bigskip

\bigskip
\noindent {\bf Connecting the ``in'' and ``out'' basis}
\bigskip

\begin{figure}[htbp]
\begin{center}
\resizebox{!}{8cm}{\input{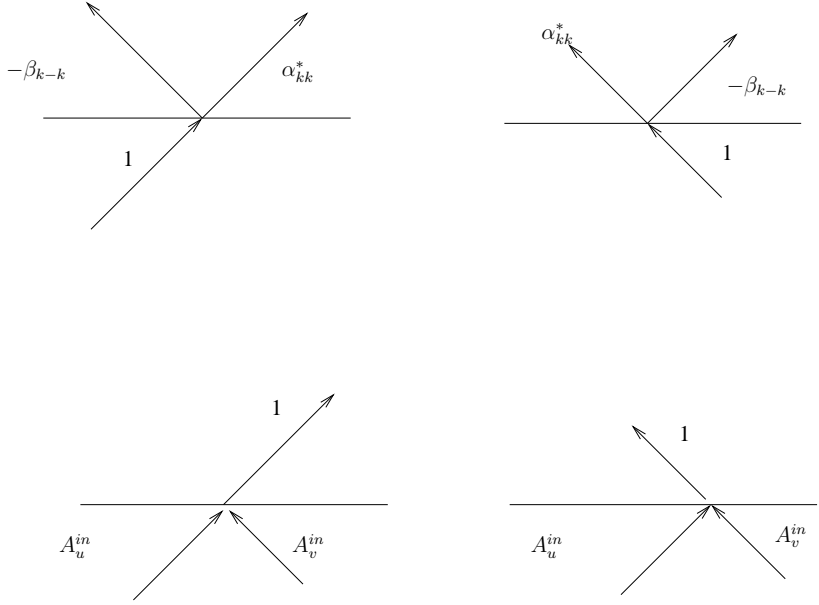}}
\caption{`in' and `out' basis in the temporal step-like discontinuity.}
\label{fig:tempgrad}
\end{center}
\end{figure}

\noindent The ``in'' and ``out'' modes basis are easily identified in terms of positive-frequency ``in'' and ``out'' modes ($u_{k,\phi}^{in(out)}=D_{ in(out)}^+(k)e^{-i\omega_+(k) t+ikx}$; for $u_{k,\varphi}^{in(out)}$ the analysis is identical up to the replacement of $D_{in(out)}^+(k)$ by $E_{ in(out)}^+(k)$) which are, respectively left-moving ($k<0$) and right moving ($k>0$). %$u_k^{in(out)}$,\ $k<0$, and right-moving $u_k^{in(out)}$,\ $k>0$.
To connect them, as depicted in Fig. \ref{fig:tempgrad}, we use the Bogoliubov matrix (\ref{eq:bogoliubovda}).
%Let us consider the evolution of right and left moving moves in the region 1 ('in' region).
%These are depicted in Fig. (\ref{fig:tempgrad}).
%\begin{figure}[htbp]
%\begin{center}
%
%\resizebox{!}{8cm}{\input{tempgrad.pstex_t}}
%
%\caption{``in'' and ``out'' modes}
%\label{figure:tempgrad}
%\end{center}
%\end{figure}
Positive frequency ``in'' modes  have amplitudes $A_{in}=1,\ B_{in}=0$. The coefficients $A_{out}$ and $B_{out}$ are found by solving the system
\begin{equation}
    \left( \begin{array}{c}
       A_{out} \\
       B_{out}  \\
     \end{array} \right)
   = \frac{1}{2\sqrt{\Omega_{in} \Omega_{out}}}\left(
            \begin{array}{cc}
               \Omega_{in}+\Omega_{out} &
               \Omega_{in}-\Omega_{out} \\
               \Omega_{in}-\Omega_{out} &
               \Omega_{in}+\Omega_{out} \\ \end{array} \right) \left(
                \begin{array}{c}
                  1  \\
                  0 \\
                \end{array}
              \right) \ ,
\end{equation}
whose solutions are
\begin{equation}
 A_{out}=\frac{\Omega_{in}+\Omega_{out}}{2\sqrt{\Omega_{in}\Omega_{out}}}\equiv \alpha_{kk}^*\ ,\qquad  B_{out}=\frac{\Omega_{in}-\Omega_{out}}{2\sqrt{\Omega_{in}\Omega_{out}}}\equiv -\beta_{k-k}\ .
 \end{equation}
These coefficients satisfy the unitarity condition
%can be easily inferred by looking at the dispersion relation plot of Fig. (\ref{figure:sub_hidro}).
%For $k>0$ we have two solutions, one (right-moving $u$ and positive frequency) belonging to the red line %(positive norm) and
%one (left-moving $u$ and negative-frequency) on the blue line (negative norm). Thus we have
\begin{equation}\label{alfabetaa} |A_{out}|^2 - |B_{out}|^2\equiv |\alpha_{kk}|^2 - |\beta_{k-k}|^2 = 1\ ,\end{equation}
%which is indeed verified by the expressions in eqs. (\ref{alfabeta}).
where the minus sign means that the $B_{out}$ is associated to negative norm modes.
% %with $|C_v^2|^2 - %|C_u^2|^2\equiv |\alpha|^2 - |\beta|^2=1$.

%\bigskip
%{\center{\bf Modes \textbf{$u_{k}^{in}\ (k>0) $}}}
%\bigskip\\
%
%A right-moving `in' mode has
%$A_v^{in}=0,\ A_u^{in}=1$. From
%
%\begin{equation}
%\label{bogu}
%     \left( \begin{array}{c}
%       A_v^{out} \\
%       A_u^{out}  \\
%     \end{array} \right)
%   = \frac{1}{2\sqrt{\Omega_{in} \Omega_{out}}}\left(
%            \begin{array}{cc}
%               \Omega_{in}+\Omega_{out} &
%               \Omega_{in}-\Omega_{out} \\
%               \Omega_{in}-\Omega_{out} &
%               \Omega_{in}+\Omega_{out} \\ \end{array} \right) \left(
%                \begin{array}{c}
%                  0  \\
%                  1 \\
%                \end{array}
%              \right) \
%\end{equation}
%  we get \begin{equation} \label{alfabetab} A_v^{out}=\frac{\Omega_{in}-\Omega_{out}}{2\sqrt{\Omega_{in}\Omega_{out}}}= -\beta_{-kk},\ \  A_u^{out}=\frac{\Omega_{in}+\Omega_{out}}{2\sqrt{\Omega_{in}\Omega_{out}}}= \alpha_{kk}^*\ .\end{equation}

%These processes describe the creation of pairs of phonons,  each pair consisting of a right-moving $u$ ($k>0$) %mode and a left-moving $v$ ($k<0$) mode with opposite momentum.
%Were the region 2 supersonic, both $u$ and $v$ modes would propagate to the left.

%We can construct the `out' modes similarly, as shown in Fig. \ref{fig:tempgrad} and developed below.

%\bigskip
%
%{\center{\bf Modes \textbf{$u_{k}^{out}\ (k<0) $}}}
%\bigskip\\

Positive frequency ``out'' modes are characterized by $A_{out}=1,\ B_{out}=0$. The coefficients $A_{in}$ and $B_{in}$ are found by solving the system
\begin{equation}
    \left( \begin{array}{c}
       1 \\
      0  \\
     \end{array} \right)
   = \frac{1}{2\sqrt{\Omega_{in} \Omega_{out}}}\left(
            \begin{array}{cc}
               \Omega_{in}+\Omega_{out} &
               \Omega_{in}-\Omega_{out} \\
               \Omega_{in}-\Omega_{out} &
               \Omega_{in}+\Omega_{out} \\ \end{array} \right) \left(
                \begin{array}{c}
                  A_{in}  \\
                  B_{in} \\
                \end{array}
              \right) \ ,
\end{equation}
 which gives
 \bea
 A_{in}=\frac{\Omega_{in}+\Omega_{out}}{2\sqrt{\Omega_{in}\Omega_{out}}}\ ,\qquad B_{in}=-\frac{\Omega_{in}-\Omega_{out}}{2\sqrt{\Omega_{in}\Omega_{out}}}\ .
 \eea

%
%\bigskip
%
%{\center{\bf Modes \textbf{$u_{k}^{out}\ (k>0) $}}}
%\bigskip\\
%A right moving 'out' mode  has $A_v^{out}=0,\ A_u^{out}=1$. The coefficients $A_v^{in}$ and $A_u^{in}$ are found by solving
%\begin{equation}
%\label{bogv}
%     \left( \begin{array}{c}
%       0 \\
%      1 \\
%     \end{array} \right)
%   = \frac{1}{2\sqrt{\Omega_{in} \Omega_{out}}}\left(
%            \begin{array}{cc}
%               \Omega_{in}+\Omega_{out} &
%               \Omega_{in}-\Omega_{out} \\
%               \Omega_{in}-\Omega_{out} &
%               \Omega_{in}+\Omega_{out} \\ \end{array} \right) \left(
%                \begin{array}{c}
%                  A_v^{in}  \\
%                  A_u^{in} \\
%                \end{array}
%              \right) \ .
%\end{equation}
%In this case we obtain $A_v^{in}=-\frac{\Omega_{in}-\Omega_{out}}{2\sqrt{\Omega_{in}\Omega_{out}}},\, A_u^{in}=\frac{\Omega_{in}+\Omega_{out}}{2\sqrt{\Omega_{in}\Omega_{out}}}$.\\
\noindent From these results, we see that the ``in'' and the  ``out'' modes are related by the relations
\begin{equation}
u_k^{in}=\alpha_{kk}^*u_k^{out}-\beta_{k-k} u_{-k}^{out*},
\end{equation}
and, considering the ``in'' and ``out'' decompositions of the field $\hat \phi$
\begin{equation}\label{phia}
\hat \phi(t,x)^{in(out)} = \int_{-\infty}^{\infty} dk\left[\hat a_{k}^{in(out)}u_{k,\phi}^{in(out)}+a_{k}^{in(out)\dagger}u_{k,\varphi}^{in(out)*}\right]\ ,
\end{equation}
we find the relation between the ``in'' and ``out'' set of operators, namely
\begin{equation}\label{hoyya}
  \hat a_{k}^{out}=\alpha_{kk}^*\hat a_k^{in}-\beta_{k-k}^*\hat a_{-k}^{in\dagger}\ .
\end{equation}
The fact that both anhilitation and creation operators enter in the r.h.s. of the above equation means that the two decompositions (\ref{phia}) are inequivalent and that $| in\rangle \neq | out\rangle$.
%We therefore obtain the following density-density:
\bigskip

\bigskip
{\center{\bf Density-density correlations}}
\bigskip

\noindent The analysis of the density-density correlation is similar to the one performed in the  hydrodynamic case, see \cite{nostro}. We first write down the operator $\hat n^1$ in the ``out'' decomposition
\begin{equation}
\hat n^1(t,x) = n\int_{-\infty}^{\infty} dk\left[a_{k}^{out}(u_{k,\phi}^{out}+u_{k,\varphi}^{out})+a_{k}^{out\dagger}(u_{k,\phi}^{out*}+u_{k,\varphi}^{out*})\right]\ ,
\end{equation}
and then use relation (\ref{hoyya}). For the two point function of $\hat n^1$ in the $|\rm in\rangle$ state we have
\begin{eqnarray}\label{fff}\non
&&\langle {\rm in}|\{ \hat n^1 (t,x), \hat n^1 (t',x')\}|{\rm in}\rangle=n^2\int_{-\infty}^{\infty}dk\left\{\left[\alpha_{kk}^*(u_{k,\phi}^{out}+u_{k,\varphi}^{out})-\beta_{k-k}(u_{-k,\phi}^{out*}+u_{-k,\varphi}^{out*})\right](t,x)\right.\times\nonumber\\
&&\left.\times\left[\alpha_{kk}(u_{k,\phi}^{out*}+u_{k,\varphi}^{out*})-\beta_{k-k}^*(u_{-k,\phi}^{out}+u_{-k,\varphi}^{out})\right](t',x')+{\rm c.c.}\right\}
\end{eqnarray}
This integral is well approximated by its hydrodynamical limit and the features of the density-density correlations are discussed in \cite{ftemp} and \cite{nostro}. %by considering the small $\omega$ approximation, and in particular for $t=t'$ it is well approximated by the hydrodynamic value
%
%
%\begin{eqnarray}\label{gtwoto}
%&G^{(2)}(t;x,x')\sim&\\
%&\frac{\hbar}{4\pi nmc_{out}}\left\{\frac{(c_{in}^2-c_{out}^2)}{2c_{in}c_{out}} \left[ \frac{1}{(2c_{out}t -(x-x'))^2} + \frac{1}{(2c_{out}t -(x'-x))^2}\right]
%- \frac{(c_{in}^2+c_{out}^2)}{c_{in}c_{out}} \frac{1}{(x-x')^2}
%\right\}\ .&\nonumber
%\end{eqnarray}
%
%At $t=0$ and everywhere in space correlated pairs of particles with opposite momentum are created out of the vacuum state,
%with velocities $v-c_{out}$ (left-moving) and $v+c_{out}$ (right-moving). At time $t$ such particles are separated by a distance $|x-x'|=2c_{out}t$, which is indeed the correlation displayed in (\ref{gtwoto}).

\subsection{Subsonic-supersonic configurations}\label{temporal_sup}

\begin{figure}[htbp]
\begin{center}

\resizebox{!}{8.5cm}{\input{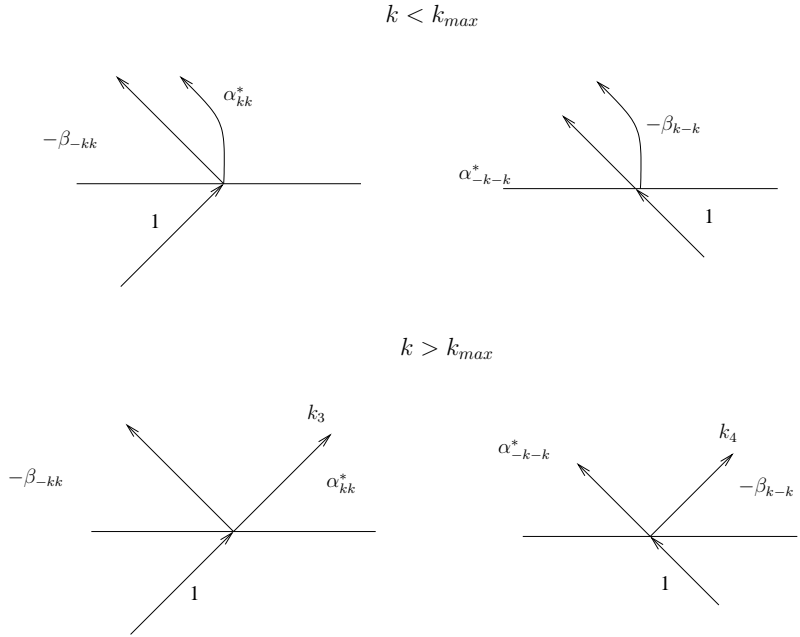}}

\caption{Evolution of `in' modes for different values of $k$ in the case of a supersonic `out' region.}
\label{fig:tempgrad2}
\end{center}
\end{figure}

\noindent This case, which is relevant  for the calculation of  section \ref{correlations}, consists in a configuration made of an ``in'' subsonic region and an ``out'' supersonic one ($c_{in}>|v|,\,\,\, c_{out}<|v|$). In the ``in'' region the analysis is the same as in the previous subsection. In the ``out'' (supersonic) one the dispersion relation (\ref{nrela}) shows new features with respect to the analysis in the hydrodynamic limit. From Fig.\  \ref{fig:tempgrad2} we see that, for $|k|<|k_{max}|$, the analysis is similar to that of the previous subsection, with the important difference that both modes are dragged by the flow and move to the left, whereas, when $|k|>|k_{max}|$, the supersonic modes $k_3(>0)$ and $k_4(<0)$ (in the language of subsection \ref{supconf})  become able to propagate to the right upstream (from now on we find more convenient to work with positive $k$, and indicate negative $k$ with $-k$). The way in which the ``in'' modes propagate in the ``out'' region is shown in Fig. \ref{fig:tempgrad2}. These features become very important for the analysis of the temporal formation of acoustic black holes of section \ref{correlations}.
%
%\bigskip
%{\center{\bf Connecting the `in' and `out' basis}}
%\bigskip

%The `in' and `out' basis are easily identified.
%They are constructed in terms of `in' and `out' positive-frequency modes which are left-moving $u_k^{in(out)}$,\ $k<0$, and right-moving $u_k^{in(out)}$,\ $k>0$.
%To connect them, as depicted in Fig. \ref{fig:tempgrad2}, we use the Bogoliubov matrix (\ref{eq:bogoliubovdd}). Were the `out' region supersonic, both the modes would move to the left.
%

%\bigskip
%{\center{\bf Modes \textbf{$u_{k}^{in}\ (k<0) $}}}
%\bigskip\\
For $k>k_{max}$, an initial left-moving mode decomposes into a positive norm left-moving component plus a $k_4$ negative norm one, with amplitudes $A_{out}$ and $A_4^{out}$ respectively. These are found by solving
%A left moving 'in' mode  has $A_v^{in}=1,\ A_u^{in}=0$. We can see from Fig. \ref{fig:tempgrad2} that the `out' modes in this case correspond to $k_v$ and $k_4$. The coefficients $A_v^{out}$ and $A_4^{out}$ are found by solving
\begin{equation}
     \left( \begin{array}{c}
       A_{out} \\
       A_4^{out}  \\
     \end{array} \right)
   = \frac{1}{2\sqrt{\Omega_{in} \Omega_{out}}}\left(
            \begin{array}{cc}
               \Omega_{in}+\Omega_{out} &
               \Omega_{in}-\Omega_{out} \\
               \Omega_{in}-\Omega_{out} &
               \Omega_{in}+\Omega_{out} \\ \end{array} \right) \left(
                \begin{array}{c}
                  1  \\
                  0 \\
                \end{array}
              \right) \ ,
\end{equation}
which yields the solutions
 \begin{equation}\label{alfabeta1}
  A_{out}=\frac{\Omega_{in}+\Omega_{out}}{2\sqrt{\Omega_{in}\Omega_{out}}}\equiv \alpha_{-k-k}^*\ ,\qquad  A_4^{out}=\frac{\Omega_{in}-\Omega_{out}}{2\sqrt{\Omega_{in}\Omega_{out}}}\equiv -\beta_{k-k}\ .
 \end{equation}
These satisfy the unitarity condition
\begin{equation}
|A_{out}|^2 - |A_4^{out}|^2\equiv |\alpha_{-k-k}|^2 - |\beta_{k-k}|^2 = 1\ .
\end{equation}
%which is indeed verified by the expressions in eqs. (\ref{alfabeta}).
% %with $|C_v^2|^2 - %|C_u^2|^2\equiv |\alpha|^2 - |\beta|^2=1$.
%\bigskip
%{\center{\bf Modes \textbf{$u_{k}^{in}\ (k>0) $}}}
%\bigskip\\
An initial right-moving mode splits instead into a positive norm right moving $k_3$ mode, with amplitude $A_3^{out}$ plus a negative norm left moving one $A_{out}$, which are found by solving
%A right-moving `in' mode has
%$A_v^{in}=0,\ A_u^{in}=1$. From Fig. \ref{fig:tempgrad2}, we see that now $k_u$ and $k_3$ are the momenta for the `out' modes. Their amplitudes can be found by solving
\begin{equation}
\label{bogu}
     \left( \begin{array}{c}
       A_{out} \\
       A_3^{out}  \\
     \end{array} \right)
   = \frac{1}{2\sqrt{\Omega_{in} \Omega_{out}}}\left(
            \begin{array}{cc}
               \Omega_{in}+\Omega_{out} &
               \Omega_{in}-\Omega_{out} \\
               \Omega_{in}-\Omega_{out} &
               \Omega_{in}+\Omega_{out} \\ \end{array} \right) \left(
                \begin{array}{c}
                  0  \\
                  1 \\
                \end{array}
              \right) \ .
\end{equation}
the solutions are
 \begin{equation} \label{alfabeta}
 A_3^{out}=\frac{\Omega_{in}-\Omega_{out}}{2\sqrt{\Omega_{in}\Omega_{out}}}= -\beta_{-kk}\ , \qquad  A_{out}=\frac{\Omega_{in}+\Omega_{out}}{2\sqrt{\Omega_{in}\Omega_{out}}}= \alpha_{kk}^*\ .
\end{equation}
Eqs. (\ref{alfabeta1}) and (\ref{alfabeta}) are the crucial formulas that we shall need in the next section to consider the temporal formation of acoustic black hole-like configurations.

\section{Density-density correlations in the formation of acoustic black hole-like configurations}\label{correlations}

\begin{figure}[htbp]
\begin{center}
\resizebox{!}{6cm}{\input{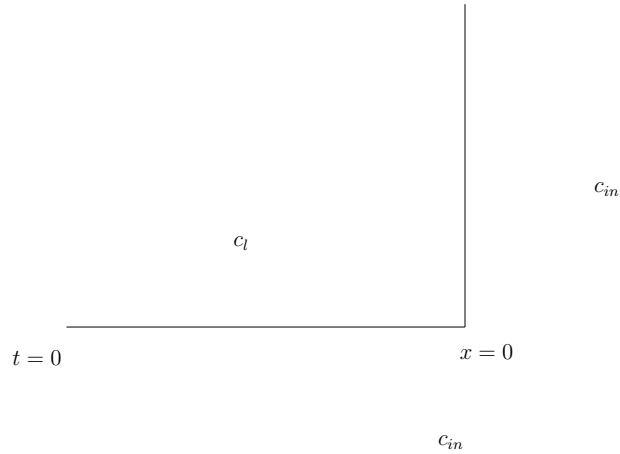}}
\caption{Temporal formation of a spatial step-like discontinuity temporally formed ($c_r=c_{in}$).}
\label{figure:crcu}
\end{center}
\end{figure}

\noindent In this section, with the help of the thorough analysis of the previous two sections, we will study the main Hawking signal in the more involved situation where an initial homogeneous subsonic flow turns supersonic in some region. We will model this situation with a temporal step-like discontinuity at $t=0$ (temporal formation) followed by a spatial step-like discontinuity at $x=0$ separating a subsonic and a supersonic region. The model we shall consider is sketched in Fig.\  \ref{figure:crcu}, where $c_r=c_{in}$.
To study the propagation of modes solutions to Eqs. (\ref{cde}) for all $x$ and $t$ we need to impose matching conditions (\ref{matching}) at $t=0$ at fixed $k$ (only those for $x<0$ are non trivial), and then (\ref{matchingaa}) at fixed $\omega$ at $x=0$ (and $t>0$). The behaviour at $x=t=0$ is more delicate because it depends the way we approach it. A detailed analysis of what happens for the case of subsonic flows was carried out in \cite{nostro} by explicitly constructing the `in' modes basis.
As modes transiting through the origin only affect transient behaviours in the correlations patterns, in this section we will rather focus on those modes solutions which give the leading contribution to the main Hawking signal. We saw in the stationary analysis of section III.B that this is given by the evolution of the modes $u_w^{3,in},\ u_w^{4,in*}$ for $w$ small and, consequently, $|k|\gtrsim |k_{max}|$. In turn, as shown in Fig. \ref{fig:tempgrad2} such modes are generated by `in' modes in the homogeneous $t<0$ region with the same value of $k$ crossing the temporal step-like discontinuity on the $x<0$ side.

In our analysis we shall need to consider a transition  from the $k$ to the $\omega$ basis. The relations between modes and operators in the two basis are
\begin{equation}\label{kw}
u_{\omega,\phi(\varphi)}=\sqrt{\frac{d\omega}{dk}}u_{k,\phi(\varphi)}\ ,\qquad\hat a_{\omega}=\sqrt{\frac{dk}{d\omega}}\hat a_{k}\ .
\end{equation}

To construct the two point function $\langle {\rm in}|\hat n^1(t,x)\hat n^1(t',x')|{\rm in}\rangle$ we proceed as usual by decomposing $\hat n^1$ in the ``out'' $\omega$ basis
\begin{eqnarray}
\hat n^1(t,x)=n\int_{0}^{\omega_{max}}\left[\hat a_{\omega}^{v,out}(u_{\omega,\phi}^{v,out}+u_{\omega,\varphi}^{v,out})+
\hat a_{\omega}^{ur,out}(u_{\omega,\phi}^{ur,out}+u_{\omega,\varphi}^{ur,out})+\hat a_{\omega}^{ul,out}(u_{\omega,\phi}^{ul,out}+u_{\omega,\varphi}^{ul,out})+{\rm h.c.}\right]\ ,
\end{eqnarray}
and by relating the $\hat a_{\omega}^{out},\hat a_{\omega}^{out\dagger}$ operators to the $\hat a_k^{in},\hat a_k^{in\dagger}$ in the ``in'' ($t<0$) region. This is done in two steps. First, the analysis in subsection \ref{supconf} provides for the relation between ``out'' and ``in'' $\omega$ basis in the $t>0$ region. In particular we have
%In section 4 we have solved the problem for $t>0$ and in particular we have that the relation between the `in' and `out' modes is
%\begin{equation}\label{inout}
%u_k^{in}=\alpha_{kk}^*u_k^{out}-\beta_{k-k} u_{-k}^{out*}.
%\end{equation}

%We compute now the density-density function for the spatial gradino temporally formed, whose set up is depicted in Fig \ref{figure:crcu}, where $c_l<|v|$ and $ c_r=c_{in}>|v|$.\\
%The first step is as in the purely spatial gradino, namely we write the $a_{\omega}^{out}$ operators in terms of the $a_{\omega}^{in}$:
%
\begin{eqnarray}
 \hat a_{\omega}^{v,out} &=& A_v^l\hat a_{\omega}^{v,in}+A_v^L\hat a_{\omega}^{3in}+A_v^{l'}\hat a_{\omega}^{4in\dagger} \ ,\\
  \hat a_{\omega}^{ur,out} &=& A_u^r\hat a_{\omega}^{v,in}+A_u^R\hat a_{\omega}^{3in}+A_u^{r'}\hat a_{\omega}^{4in\dagger}\ , \\
  \hat a_{\omega}^{ul,out\dagger} &=& A_u^l\hat a_{\omega}^{v,in}+A_u^L\hat a_{\omega}^{3in}+A_u^{l'}\hat a_{\omega}^{4in\dagger}\ .
\end{eqnarray}
From the values of the amplitudes in the above equation (given in subsection \ref{supconf} ) we see that the terms multiplying  $\hat a_{\omega}^{v,in}$ are  subleading with respect to those multiplying $\hat a_{\omega}^{3in}$ and $\hat a_{\omega}^{4in\dagger}$.

Next, we need to jump from the ``in'' $\omega$-basis to the $k$-basis needed to address the temporal step-like discontinuity. The relevant terms in $\hat n^1$ in our analysis are
\begin{equation}
\hat n^1(t,x) = \int_{k_{max}}^{\infty} dk_3\left[\hat a_{k_3}(u_{k_3,\phi}+u_{k_3,\varphi})+\hat a_{k_4}^{\dagger}(u_{k_4,\phi}^*+u_{k_4,\varphi}^*)\right],
\end{equation}
where $k_4=-k_3$. This is to be matched, at $t=0$, at the relevant values of $k$, with the ``in'' decomposition ($t<0$)
\begin{equation}
\hat n^1(t,x) = \int_{0}^{\infty} dk\left[\hat a_{k}^{in}(u_{k,\phi}^{in}+u_{k,\varphi}^{in})+\hat a_{-k}^{in}(u_{-k,\phi}^{in}+u_{-k,\varphi}^{in})+\hat a_{k}^{in\dagger}(u_{k,\phi}^{in*}+u_{k,\varphi}^{in*})+\hat a_{-k}^{in\dagger}(u_{-k,\phi}^{in*}+u_{-k,\varphi}^{in*})\right].
\end{equation}
The relations between the $k$ operators before and after the temporal step-like discontinuity are given by (\ref{hoyya}) with $k_4=-k_3$:
\begin{eqnarray}\label{hoyy}
  \hat a_{k_3} &=& \alpha(k_3)\hat a_k-\beta^*(-k_3)\hat a_{-k}^{\dagger}\ , \nonumber\\
  \hat a_{-k_3}^\dagger &=& -\beta(-k_3)\hat a_{-k}+\alpha^*(k_3)\hat a_k^\dagger\ ,
\end{eqnarray}
where the Bogoliubov coefficients are given by (\ref{alfabeta1}) and (\ref{alfabeta})
\begin{eqnarray}\label{bogob}
   \alpha=\frac{\Omega_{in}+\Omega_{out}}{2\sqrt{\Omega_{in}\Omega_{out}}}\ ,\qquad   \beta=\frac{\Omega_{out}-\Omega_{in}}{2\sqrt{\Omega_{in}\Omega_{out}}}\ .
\end{eqnarray}
Here $\Omega=|\omega-vk|$ is calculated before ($\Omega_{in}$) and after ($\Omega_{out}$) the temporal discontinuity.

\begin{figure}[htbp]
\begin{center}
\resizebox{!}{5cm}{\input{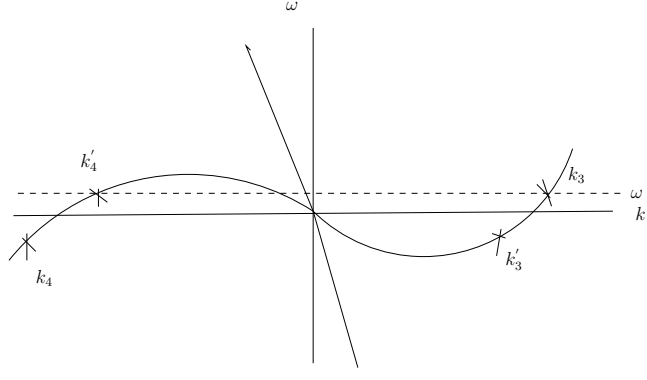}}
\caption{$\omega$ versus $k$ in the supersonic case.}
\label{figure:temporalfor}
\end{center}
\end{figure}

Let us now go back to the $\omega$ basis (the general relation between modes and operators in the $\omega$ and $k$ basis is given in (\ref{kw})). As shown in Fig.\  \ref{figure:temporalfor}, a fixed, positive value of $\omega$ corresponds to two values of $k$, namely $k_3$ and $k_4'$. We thus write
\begin{equation}
\hat n^1(t,x) = \int_{0}^{\omega_{max}} d\omega\left[\hat a_{\omega}^{3in}(u_{\omega,\phi}^{3in}+u_{\omega,\varphi}^{3in})+\hat a_{\omega}^{4in\dagger}(u_{\omega,\phi}^{4in*}+u_{\omega,\varphi}^{4in*})+h.c.\right].
\end{equation}
Defining $k_4=-k_3$ and $k_4'=-k_3'$, the following properties are valid (we do not write explicitly the normalizations)
\begin{equation}
u_{\omega}^{3in*}=\left(e^{-i\omega t+ik_3(\omega)x}\right)^*=e^{i\omega t-ik_3(\omega)x}=e^{-i(-\omega)t+i(-k_3(\omega))x}=e^{-i(-\omega)t+ik_4(-\omega)x}=u_{-\omega}^{4in*}
\end{equation}
and
\begin{equation}
u_{\omega}^{4in*}=e^{-i\omega t+ik_4'(\omega)x}=e^{-i(-\omega)t+i(-k_3'(-\omega))x}=u_{-\omega}^{3in*}.
\end{equation}
Therefore the density fluctuation operator $\hat n^1$ turns into:
\begin{eqnarray}
\hat n^1(t,x) = \int_{0}^{\omega_{max}}\!\!\!\!d\omega\left[\hat a_{\omega}^{3in}(u_{\omega,\phi}^{3in}+u_{\omega,\varphi}^{3in})+\hat a_{\omega}^{4in\dagger}(u_{\omega,\phi}^{4in*}+u_{\omega,\varphi}^{4in*})+\hat a_{-\omega}^{4in\dagger}(u_{-\omega,\phi}^{4in*}+u_{-\omega,\varphi}^{4in*})+\hat a_{-\omega}^{3in}(u_{-\omega,\phi}^{3in}+u_{-\omega,\varphi}^{3in})\right].
\end{eqnarray}
Since the $\omega$ decomposition requires two values of $k$, the relation between the $\omega$ operators and the $k$ ones before the temporal discontinuity will involve relations (\ref{hoyy}) with different values of $k$, namely $k_3\equiv k$ and $k_4'\equiv -k'$:
\begin{eqnarray}
  \hat a_{\omega}^{3in} &=& \alpha(k_3)\sqrt{\frac{d\omega}{dk_3}}\hat a_{k_3}-\beta^*(-k_3)\sqrt{\frac{d\omega}{dk_3}}\hat a_{-k_3}^{\dagger}\ , \\
  \hat a_{\omega}^{4in\dagger} &=& -\beta(-k_3'(-\omega))\sqrt{\frac{d\omega}{dk_3'}}\hat a_{-k'}+\alpha^*(k_3'(-\omega))\sqrt{\frac{d\omega}{dk_3'}}\hat a_{k'}^{\dagger}\ .
\end{eqnarray}
We compute now the Bogoliubov coefficients appearing above. Let us start with $\alpha(k_3)$ and $\beta(-k_3)$.
%\begin{eqnarray}\label{bogob}
%   \alpha(k_3)&=&\frac{|\Omega_1|+|\Omega_2|}{2\sqrt{|\Omega_1\Omega_2|}},\nonumber\\
%   \beta(-k_3)&=&\frac{|\Omega_2|-|\Omega_1|}{2\sqrt{|\Omega_1\Omega_2|}}.
%\end{eqnarray}
%We can also use the small $\omega$ expression for $k_3$, while for $k$ we have to use the full dispersion relation:
%\begin{eqnarray}
%% \nonumber to remove numbering (before each equation)
%  k_3 &=& \frac{2\sqrt{v^2-c_l^2}}{c_l\xi_l}+\frac{v\omega}{c_l^2-v^2} \\
%  k &=& c_{in}\sqrt{k^2+\frac{k^4\xi_{in}^2}{4}}.
%\end{eqnarray}
By using again the fact that $k$ is conserved in the temporal step-like discontinuity ($k=k_3$) and the expression of $k_3$ for small $\omega$ ($k_3 = \frac{2\sqrt{v^2-c_l^2}}{c_l\xi_l}+\frac{v\omega}{c_l^2-v^2}$), which gives the main contribution to the density-density correlations, we can write $\Omega_{in}$ and $\Omega_{out´}$ as:
\begin{eqnarray}
  \Omega_{out} &=& \omega-v\left[\frac{2\sqrt{v^2-c_l^2}}{c_l\xi_l}+\frac{v\omega}{c_l^2-v^2}\right]\ , \\
  \Omega_{in} &=& c_{in}\sqrt{k^2+\frac{k^4\xi_{in}^2}{4}}\ .
\end{eqnarray}
Notice that we cannot use the perturbative expressions in the ``in'' region, since here we are beyond the small frequency regime.
Expanding up to $\omega$ we finally obtain:
\begin{eqnarray}
   &&\alpha(k_3)=\frac{-v+\sqrt{v^2-c_l^2+c_{in}^2}}{2\sqrt{-v}(v^2-c_l^2+c_{in}^2)^{1/4}}+\frac{c_l\sqrt{-v}(c_l^2-c_{in}^2)(v+\sqrt{v^2-c_l^2+c_{in}^2})\xi_l \omega}{8v^2\sqrt{v^2-c_l^2}(v^2-c_l^2+c_{in}^2)^{5/4}}\ ,\\
   &&\beta(-k_3)=-\frac{v+\sqrt{v^2-c_l^2+c_{in}^2}}{2\sqrt{-v}(v^2-c_l^2+c_{in}^2)^{1/4}}+\frac{c_l\sqrt{-v}(c_l^2-c_{in}^2)(-v+\sqrt{v^2-c_l^2+c_{in}^2})\xi_l \omega}{8v^2\sqrt{v^2-c_l^2}(v^2-c_l^2+c_{in}^2)^{5/4}}\ .\nonumber
\end{eqnarray}
Let us compute now $\alpha(k_3')$ and $\beta(-k_3')$.
%\begin{eqnarray}\label{bogob}
%   \alpha(k_3')&=&\frac{|\Omega_1|+|\Omega_2|}{2\sqrt{|\Omega_1\Omega_2|}}\nonumber\\
%   \beta(-k_3')&=&\frac{|\Omega_2|-|\Omega_1|}{2\sqrt{|\Omega_1\Omega_2|}}
%\end{eqnarray}
%The perturbative $k$ are:
%\begin{eqnarray}
%% \nonumber to remove numbering (before each equation)
%  -k_3'(-\omega) &=& -\frac{2\sqrt{v^2-c_l^2}}{c_l\xi_l}+\frac{v\omega}{c_l^2-v^2} \\
%  -k &=& c_{in}\sqrt{k^2+\frac{k^4\xi_{in}^2}{4}}.
%\end{eqnarray}
By using the fact that $k$ is conserved in the temporal step-like discontinuity ($-k=k_4$) and the expression of $-k_3'$ for small $\omega$ ($-k_3'(-\omega) = -\frac{2\sqrt{v^2-c_l^2}}{c_l\xi_l}+\frac{v\omega}{c_l^2-v^2}$) we have
\begin{eqnarray}
  \Omega_2 &=&-\omega+v\left[-\frac{2\sqrt{v^2-c_l^2}}{c_l\xi_l}+\frac{v\omega}{c_l^2-v^2}\right]\ , \\
  \Omega_1 &=& c_{in}\sqrt{k^2+\frac{k^4\xi_{in}^2}{4}}\ .
\end{eqnarray}
By expanding (\ref{bogob}) up to $\omega$ these expressions we finally obtain, at that perturbative level,
\begin{eqnarray}
   \alpha(k_3')=\alpha(k_3)\ ,\qquad   \beta(-k_3')=\beta(-k_3)\ .
\end{eqnarray}

\begin{figure}[ht]
\centering
\includegraphics[width=70mm,height=60mm]{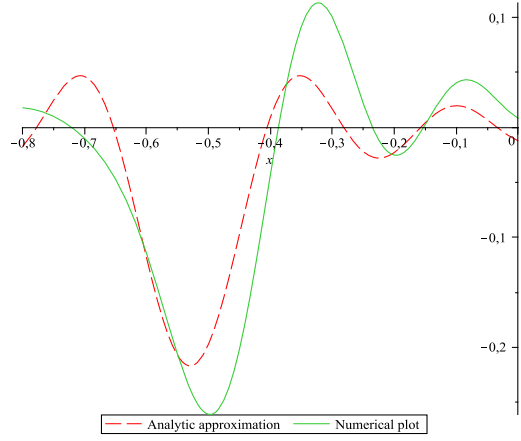}
\caption{Comparison between the plots of Eq.\ (\ref{correlation11}) and of the numerical counterpart along the direction $x=x'-1$, where $x$ is in units of the length $\xi$. We adopted the following numerical values: $v=-1.01, \quad c_l=-v/4, \quad c_r=-5v/3, \quad n=5.1, \quad m=20.1$. With these choices $\xi\simeq 0.03$ with $\hbar=1$. These values have been chosen of the same order of  the ones used in the  simulations studied in \cite{fnum}, so that they qualitatively match the results of  \cite{recatipavloff}.}
\label{tempoformed}
\end{figure}

\begin{figure}[ht]
\centering
\includegraphics[width=130mm,height=90mm]{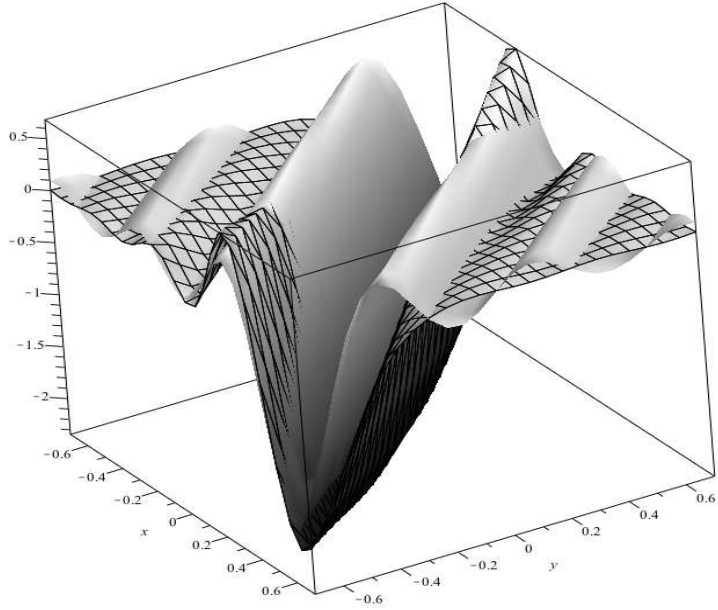}
\caption{Comparison between the plot of Eq.\ (\ref{correlation11}) and of the numerical counterpart, obtained by numerically solving the integrals without truncating the expressions for the momenta as in appendix C. The variable $y$ corresponds to $x'$. The values are as in Fig.\ (\ref{tempoformed}).
%We adopted the values $v=-1.01, \quad c_l=-v/4, \quad c_r=-5v/3 $. 
}
\label{tempoformedtd}
\end{figure}

We have now all the ingredients to calculate the main contribution to the Hawking signal in the density-density correlation for the temporally formed step. We study again the correlation between the modes $u_{\omega}^{ur}$ and $u_{\omega}^{ul*}$. As at the end of subsection \ref{supconf},  $x$ ($x'$) is a point in the left (right) region. The two-point function reads
\begin{eqnarray}\non
&&\langle {\rm in}|\{ \hat n^1 (t,x), \hat n^1 (t',x')\}|{\rm in}\rangle(u_{\omega}^{ur}\leftrightarrow u_{\omega}^{ul*})=\\
&&n^2\int_{0}^{\omega_{max}}d\omega\left[\left(A_u^{l'*}A_u^{r'}|\alpha(k_3')|^2+A_u^{R}A_u^{L*}|\beta(-k_3)|^2\right)(u_{\omega,\phi}^{ul}+u_{\omega,\varphi}^{ul})(t,x)(u_{\omega,\phi}^{ur,out}+u_{\omega,\varphi}^{ur,out})(t,x')+\right.\nonumber\\
&&\left.+\left(A_u^{L}A_u^{R*}|\alpha(k_3)|^2+A_u^{L}A_u^{R*}|\beta(-k_3')|^2\right)(u_{\omega,\phi}^{ul,out*}+u_{\omega,\varphi}^{ul*})(t,x)(u_{\omega,\phi}^{ur*}+u_{\omega,\varphi}^{ur,out*})(t,x')+\,{\rm c.c.}\right].
%\int_0^{\omega_{max}}d\omega\left[A_2^{l'*}A_2^{r'}(D+E)|_{k_2^l}(D+E)|_{k_2^r}u_{-\omega}^{2lout}(t,x)u_{\omega}^{2rout}(t,x')\right]
\end{eqnarray}
The products of the amplitudes are related by
\begin{equation}
[a_{\omega}^{ur,out},a_{\omega}^{ul,out}]=0\Rightarrow A_u^{l*}A_u^r+A_u^{L*}A_u^R-A_u^{l'*}A_u^{r'}=0,
\end{equation}
where we have used the relation between ``in'' and ``out'' $\omega$ operators.
We neglect the subleading term $A_u^{l*}A_u^r$ and take into account that at leading order $A_u^{l'*}A_u^{r'}$ is real. Thus, we find
\begin{eqnarray}\label{l}
&&\langle {\rm in}|\{ \hat n^1 (t,x), \hat n^1 (t',x')\}|{\rm in}\rangle (u_{\omega}^{ur}\leftrightarrow u_{\omega}^{ul*})=n^2\int_{0}^{\omega_{max}}d\omega \left\{\left(|\alpha(k_3)|^2+|\alpha(k_3')|^2+|\beta(-k_3)|^2+|\beta(-k_3')|^2\right)\right.\nonumber \times \\
&&\left.\times \,\,A_u^{l'*}A_u^{r'}Re\left[(u_{\omega,\phi}^{ul}+u_{\omega,\varphi}^{ul})(t,x)(u_{\omega,\phi}^{ur}+u_{\omega,\varphi}^{ur})(t,x')\right]+{\rm c.c.}\right\}.
\end{eqnarray}
%where
%\begin{equation}
%    A_u^{l'*}A_u^{r'}(D_u^l+E_u^l)(D_u^r+E_u^r)\sim \frac{1}{2\pi n}\frac{(v^2-c_l^2)^{3/2}}{2c_l(v+c_l)(v-c_r)(c_l-c_r)}.
%\end{equation}
%
%%\begin{equation}
%%    (u_{\omega,\phi}^{ul,out}+u_{\omega,\varphi}^{ul,out})(t,x)(u_{\omega,\phi}^{ur,out}+u_{\omega,\varphi}^{ur,out})(t,x')\simeq e^{i\omega \left(\frac{x}{v+c_l}-\frac{x'}{v+c_r}\right)}\ ,
%%\end{equation}
By taking in account that
\begin{equation}\label{bogo2}
|\alpha(k_3)|^2+|\alpha(k_3')|^2+|\beta(-k_3)|^2+|\beta(-k_3')|^2=\frac{c_l^2-c_{in}^2-2v^2}{v\sqrt{v^2-c_l^2+c_{in}^2}}\ ,
\end{equation}
we can finally write down the leading order contribution to $G^{(2)}$, namely
\begin{equation}\label{correlation11}
G^{(2)}(t;x,x') =\frac{1}{4\pi n}\frac{(v^2-c_l^2)^{3/2}(c_l^2-c_{r}^2-2v^2)}{2vc_l(v+c_l)(v-c_r)(c_l-c_r)\sqrt{v^2-c_l^2+c_{r}^2}}\frac{\sin\left[\omega_{max}(\frac{x'}{v+c_r}-\frac{x}{v+c_l})\right]}{\frac{x'}{v+c_r}-\frac{x}{v+c_l}}\ ,
\end{equation}
which modifies the stationary correlation (\ref{correlation_spatial}) by the factor (\ref{bogo2}) that comes from the effect of the temporal formation. In Fig.\ (\ref{tempoformed}) we display the plots of Eq.\ (\ref{correlation11}), and of the numerical counterpart along the direction $x=x'-1$. The picture shows a good agreement, which confirms that the analytic approximation adopted in this paper is good enough to capture the essential features of the correlations. Good agreement exists also for different cuts, for completeness a 3D contour plot is given in Fig. \ref{tempoformedtd}. 

\begin{figure}[ht]
\centering
\includegraphics[width=70mm,height=60mm]{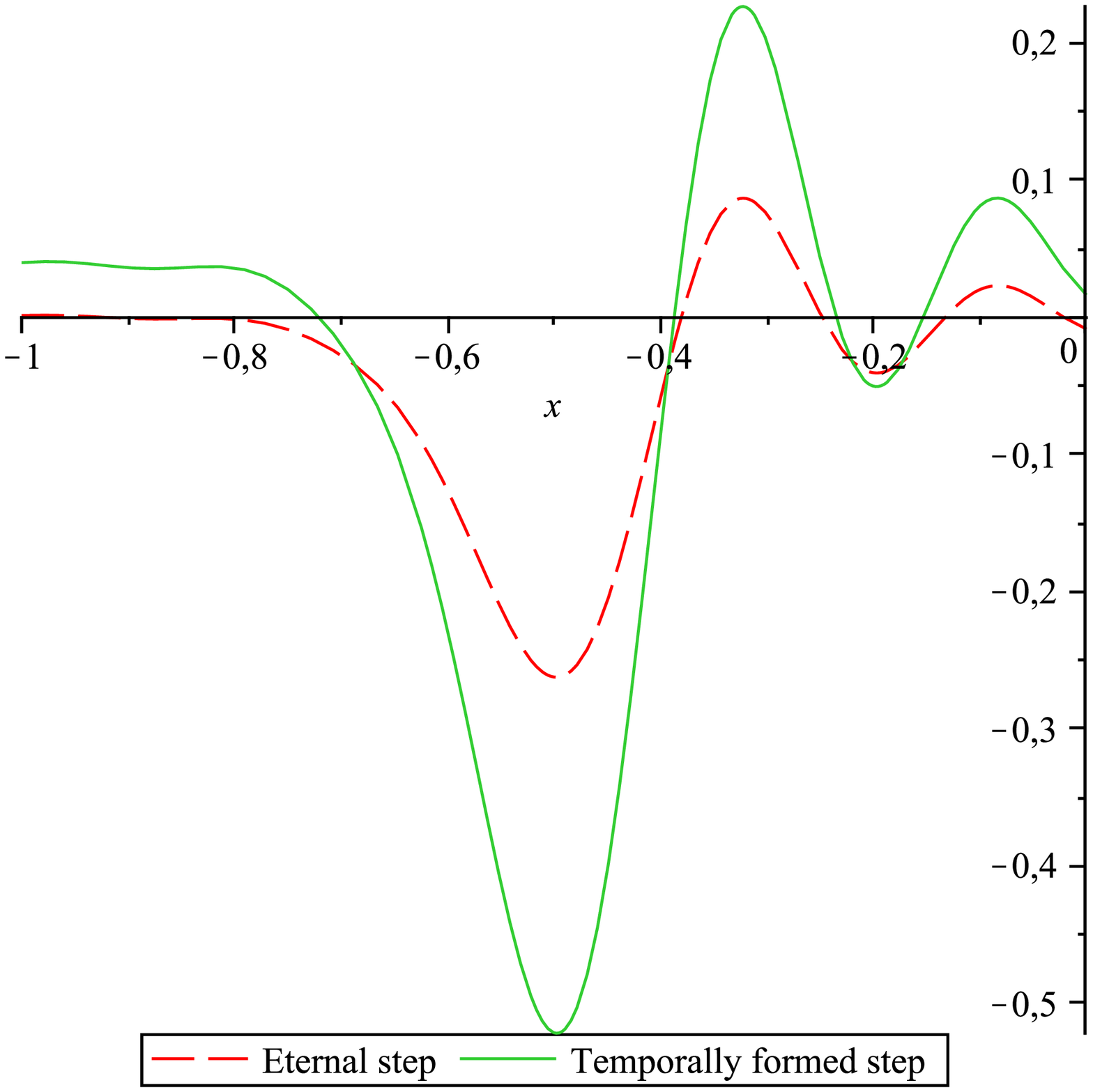}
\caption{Comparison between the (numerical) plots of the two-point function for the eternal step (\ref{correlation_spatial}) and for the temporally formed step (\ref{correlation11}) along the line  $x=x'-1$ where $x$ is in units of the length $\xi$. The choice of the parameters is the same as in Fig.\ (\ref{tempoformed}).}
\label{comparisonEtTemp}
\end{figure}

Finally, in Fig.\ (\ref{comparisonEtTemp}) we confront the signal between the eternal step and the temporally formed one. As one can see already from the analytic approximation, the temporal formation of the step yields an amplification of the signal.


\section{Final comments}\label{Final}

\noindent In this paper we have studied in detail the formation of acoustic black hole-like configurations in BECs using step-like discontinuities. The Hawking signal in the stationary case (\ref{correlation_spatial}) and in the case of temporal formation (\ref{correlation11}) have stationary peaks (at $\frac{x'}{v+c_r}=\frac{x}{v+c_l}$) of order $O(\omega_{max})\sim O(1/\xi)$, which lie well inside the non perturbative regime in $\xi$. The results in the hydrodynamical limit of \cite{fteo} showed instead a peak of order $\kappa^2$, where $\kappa=\frac{dc}{dx}|_{x=0}$ is the surface gravity of the horizon. It is clear that in the approximation of spatial step-like discontinuities we are working with the surface gravity is formally infinite and therefore our expression (\ref{correlation11}) (and also (\ref{correlation_spatial})) regularizes the result of \cite{fteo} in the $\kappa \rightarrow \infty$ limit, in agreement with the numerical results of \cite{fnum}.

In \cite{nostro}, it was noted that a simple recipe to take into account a smooth transition region in $c(x)$ around $x=0$ of width $\sigma_x$ and surface gravity $\kappa\sim \frac{c}{\sigma_x}$ is to introduce a cut-off of order $\kappa$ in the $\omega$ integral of (\ref{l}) by multiplying the integrand by the function $e^{-\omega/\kappa}$. The interplay between $\omega_{max}$ and $\kappa$ is such that the final peak is of order $\kappa(1-e^{-\omega_{max}/\kappa})$, which has the correct $\kappa >> \omega_{max}$ limit, i.e. $\omega_{max}$. However, it  is not able to make contact with the results of the hydrodymamic limit since, when $\omega_{max}>>\kappa$, we have a behavior in $\kappa$ which is linear and not quadratic. Thus, it would be interesting to find an analytical formula capable to interpolate successfully between these two limits.

\begin{acknowledgments}
\noindent We thank R.\ Balbinot, I.\ Carusotto, R.\ Parentani, and A.\ Recati for useful discussions. A.\ F.\ wishes to thank Generalitat Valenciana for financial support. A.\ F.\  and C.\ M.\  are supported in part by MICINN grant FIS2008-06078-C03-02. M.\ R.\ is supported by the Swiss National Science Foundation.
\end{acknowledgments}

%%%%%% begin appendices %%%%

\appendix

\section{}

\noindent In this appendix we construct the exact ``in'' and ``out'' basis for the spatial step-like discontinuities at $x=0$ with $v=0$ (perturbative results in $z_l=\xi \omega/c$ are given in the main text), which have a special interest for the validity of the unitarity relations. The scattering matrix determined by the junction conditions is given by \bea\label{smat}
M_{\rm scatt}=W_l^{-1}W_r\ ,
\eea
where $W_l$ and $W_r$ are given by (\ref{eq:wl}) and (\ref{eq:wr}) respectively.
%\bea
%W_l=\left(\begin{array}{cccc}{D_v^l} & {D_u^l} & G_{\phi}^l & D_{\phi}^l  \\{k_v^l}{D_v^l} & {k_u^l}{D_u^l} & {k_d^l}G_{\phi}^l & {k_g^l}D_{\phi}^l\\{E_v^l} & {E_u^l} & G_{\varphi}^l & D_{\varphi}^l \\{k_v^l}{E_v^l} & {k_u^l}{E_u^l} & {k_d^l}G_{\varphi}^l & {k_g^l}D_{\varphi}^l\end{array}\right)\ ,
%\eea
We recall that the structure of these matrices is determined uniquely by the matching conditions and by the solutions to the dispersion relation  (\ref{nrela}) on the two sides.  For $v=0$, this equation reduces to
\bea\label{masteqn}
\omega^2=c_l^2\left[(k^l)^2+{\xi_l^2(k^l)^4\over 4}\right]\ ,\quad \omega^2=c_r^2\left[(k^r)^2+{\xi_r^2(k^r)^4\over 4}\right]\ ,
\eea
where $c_{r,l}$ are the speed of sound on the right-hand side and on the left-hand side of the step respectively.
The solutions are, on the left-hand side
\bea\label{soll}
k_{u,v}^{l}&=&\pm{\sqrt{2}\over\xi_l}\sqrt{-1+\sqrt{1+{\omega^2\xi_l^2\over c_l^2}}}\ ,\\\non\\
k_{d,g}^{l}&=&\pm{i\sqrt{2}\over\xi_l}\sqrt{1+\sqrt{1+{\omega^2\xi_l^2\over c_l^2}}}\ .
\eea
Similarly, on the right-hand side, we have
\bea\label{solr}
k_{u,v}^{r}&=&\pm{\sqrt{2}\over\xi_r}\sqrt{-1+\sqrt{1+{\omega^2\xi_r^2\over c_r^2}}}\ ,\\\non\\
k_{d,g}^{r}&=&\pm{i\sqrt{2}\over\xi_r}\sqrt{1+\sqrt{1+{\omega^2\xi_r^2\over c_r^2}}}\ .
\eea
As we have $k_u=-k_v$ and $k_d=-k_g$, and hence  $D_u=D_v$  (see Eq. (\ref{eq:normdispersion})). Therefore, the matrices $W_l$ and $W_r$ simplify to
\bea\label{Wleft}
W_l=\left(\begin{array}{cccc}{D_u^l} & {D_u^l} & G_{\phi}^l & d_{\phi}^l  \\i{k_v^l}{D_u^l} & -i{k_v^l}{D_u^l} & -i{k_d^l}G_{\phi}^l & i{k_g^l}d_{\phi}^l\\{E_u^l} & {E_u^l} & G_{\varphi}^l & d_{\varphi}^l \\i{k_v^l}{E_u^l} & -i{k_v^l}{E_u^l} & -i{k_d^l}G_{\varphi}^l & i{k_g^l}d_{\varphi}^l\end{array}\right)\ ,
\eea
and
\bea\label{Wright}
W_r=\left(\begin{array}{cccc}{D_u^r} & {D_u^r} & d_{\phi}^r & G_{\phi}^r  \\i{k_v^r}{D_u^r} & -i{k_v^r}{D_u^r} & i{k_d^r}d_{\phi}^r & -i{k_g^r}G_{\phi}^r\\{E_u^r} & {E_u^r} & d_{\varphi}^r & G_{\varphi}^r \\i{k_v^r}{E_u^r} & -i{k_v^r}{E_u^r} & i{k_d^r}d_{\varphi}^r & -i{k_g^r}G_{\varphi}^r\end{array}\right)\ .
\eea
Let us now construct explicitly the `in' and `out' modes (see details in section 3.1).

\vspace{0.4cm}
\noindent{\bf $\bullet$ Mode \textbf{$u_{\omega,\phi}^{u,in}$}}
\vspace{0.4cm}

Matching conditions at the step dictates that
\bea
\left(\begin{array}{c}A_v^l \\ 1 \\0 \\A_d^l\end{array}\right)=M_{\rm scatt}\left(\begin{array}{c} 0 \\ A^r_u \\A^r_d \\0\end{array}\right)\ ,
\eea
and $R'=A_v^l$ and $T'=A_u^r$ are the reflection and transmission coefficients respectively. By solving  the system we find
\bea
|R'|^2&=&{({k_v^l}- {k_v^r})^2\over ( {k_v^l}+{k_v^r})^2}\ ,\\
|T'|^2&=&{4\left[ (k_v^l)^2-(k_d^l)^2 \right] \left[ (k_v^l)^2-(k_d^r)^2\right]\left[ (k_v^r)^2+{2\omega\over \xi_r c_r}\right]^2
 (k_v^l)^2\over
 \big[  (k_v^r)^2-(k_d^l)^2\big]\big[  (k_v^r)^2-(k_d^r)^2\big]\left[  (k_v^l)^2+{2\omega\over \xi_l c_l}\right]^2\big[  {k_v^l}+ {k_v^r}\big]^2}
 \left| {(D_1^l)^2\over (D_1^r)^2}
 \right| \ .
\eea
Note that both $R'$ and $T'$ do not depend on the normalizations $d_{\phi(\varphi)}^{l,r}$.
To simplify the above expressions, we first use the definitions of ${D_u^l}$ and ${D_u^r}$ displayed in (\ref{eq:normdispersion}) to find
\bea
|T'|^2={4(k_v^r)^2\big[(k_v^l)^2-(k_d^l)^2\big]\big[(k_v^l)^2-(k_d^r)^2\big]\over \big[(k_v^r)^2-(k_d^r)^2\big]\big[(k_v^r)^2-(k_d^l)^2\big]\big[{k_v^l}+{k_v^r}\big]^2}\left|d{k_v^l}\over d{k_v^r} \right|\ .
\eea
By calculating explicitly $d{k_v^l}/d{k_v^r}$ and by using the identities
\bea
\sqrt{\omega^2\xi_l^2+c_l^2}=c_l\left({\xi_l(k^l_v)^2\over 2}+1\right)\ ,\quad \sqrt{\omega^2\xi_r^2+c_r^2}=c_r\left({\xi_r(k^r_v)^2\over 2}+1\right)\ ,
\eea
we find
\bea
{d{k_v^l}\over d{k_v^r}}={\xi_l^2k_v^r[(\xi_rk_v^r)^2+2]\over \xi_r^2k_v^l[(\xi_lk_v^l)^2+2]}\ .
\eea
By noting further that
\bea
(k_v^l)^2-(k_d^l)^2={4\over c_l\xi_l^2}\sqrt{\omega^2\xi_l^2+1}\ ,\quad (k_v^r)^2-(k_d^r)^2={4\over c_r\xi_r^2}\sqrt{\omega^2\xi_r^2+1}\ ,
\eea
we can write $|T'|^2$ as
\bea
|T'|^2={4(k_v^r)^3\left[(k_v^l)^2-(k_d^r)^2\right]\over {k_v^l}\big[(k_v^r)^2-(k_d^l)^2\big]\big[{k_v^l}+{k_v^r}\big]^2}\ .
\eea
Finally, with the relations
\bea
(k_d^r)^2=-{4\omega^2\over (c_r\xi_rk_v^r)^2}\ ,\quad (k_d^l)^2=-{4\omega^2\over (c_l\xi_lk_v^l)^2}\ ,
\eea
which can be easily proved with Eqs. (\ref{masteqn}) and either (\ref{soll}) or (\ref{solr}), we find that
\bea
|T'|^2={4{k_v^r}{k_v^l}\over ({k_v^l}+{k_v^r})^2}\ .
\eea
Thus, $|T'|^2+|R'|^2=1$.

\vspace{0.4cm}
\noindent{\bf $\bullet$ Mode \textbf{$u_{\omega,\phi}^{v,in}$}}
\vspace{0.4cm}

The scattering matrix is still given by (\ref{smat}), but now the system to solve is

\bea
\left(\begin{array}{c}A_v^l \\ 0 \\0 \\A_d^l\end{array}\right)=M_{\rm scatt}\left(\begin{array}{c} 1 \\ A^r_u \\A^r_d \\0\end{array}\right)\ ,
\eea
with $R=A_u^r$ and $T=A_v^l$. Despite the fact that the system is different, we find the same results as in the previous case, namely
\bea
|R|^2&=&{({k_v^l}- {k_v^r})^2\over ( {k_v^l}+{k_v^r})^2}\ ,\\
|T|^2&=&{4\big[ (k_v^r)^2-(k_d^r)^2 \big]\big[ (k_v^r)^2-(k_d^l)^2\big]\left[ (k_v^l)^2+{2\omega\over \xi_l c_l}\right]^2
 (k_v^r)^2\over
 \big[ (k_v^l)^2-(k_d^r)^2\big]\big[ (k_v^l)^2-(k_d^l)^2\big]\left[ (k_v^r)^2+{2\omega\over \xi_r c_r}\right]^2\big[ {k_v^l}+ {k_v^r}\big]^2}
 \left| {(D_u^r)^2\over (D_u^l)^2}
 \right| \ ,
\eea
thus even in this case unitarity holds.
It is easy to see that these expressions can be obtained by the ones in the case $u^{u,in}_\omega$ by swapping $r \leftrightarrow l$, so that the proof that $|R|^2+|T|^2=1$ follows immediately. The construction of $u^{u,out}_{\omega,\phi}$ and $u^{v,out}_{\omega,\phi}$ is now straightforward.

\newpage

\vspace{0.4cm}
\noindent{\bf $\bullet$ Mode \textbf{$u_{\omega,\phi}^{u,out}$}}
\vspace{0.4cm}

\noindent For \textbf{$u_{\omega,\phi}^{u,out}$} we solve
\bea
\left(\begin{array}{c}0 \\ A_u^l \\0 \\A_d^l\end{array}\right)=M_{\rm scatt}\left(\begin{array}{c} A^r_v \\ 1 \\A^r_d \\0\end{array}\right)\ ,
\eea
where we identify $R^*=A_1^r$ and $T^*=A_2^l$ and find
\bea
|R^*|^2&=&{({k_v^l}- {k_v^r})^2\over ( {k_v^l}+{k_v^r})^2}\ ,\\
|T^*|^2&=&{4\big[ (k_v^r)^2-(k_d^r)^2 \big]\big[ (k_v^r)^2-(k_d^l)^2\big]\left[ (k_v^l)^2+{2\omega\over \xi_l c_l}\right]^2
 (k_v^r)^2\over
 \big[ (k_v^l)^2-(k_d^r)^2\big]\big[ (k_v^l)^2-(k_d^l)^2\big]\left[ (k_v^r)^2+{2\omega\over \xi_r c_r}\right]^2\big[ {k_v^l}+ {k_v^r}\big]^2}
 \left| {(D_u^r)^2\over (D_u^l)^2}
 \right| \ .
\eea
\vspace{0.4cm}
\noindent{\bf $\bullet$ Mode \textbf{$u_{\omega,\phi}^{v,out}$}}
\vspace{0.4cm}

For \textbf{$u_{\omega,\phi}^{v,out}$} we solve
\bea
\left(\begin{array}{c}1 \\ A_u^l \\0 \\A_d^l\end{array}\right)=S\left(\begin{array}{c} A_v^r \\ 0 \\A_d^r \\0\end{array}\right)\ ,
\eea
where $R'^*=A_u^l$ and $T'^*=A_v^r$ and get
\bea
|R'^*|^2&=&{({k_v^l}- {k_v^r})^2\over ( {k_v^l}+{k_v^r})^2}\ ,\\\non\\
|T'^*|^2&=&{4\left[ (k_v^l)^2-(k_d^l)^2 \right] \left[ (k_v^l)^2-(k_d^r)^2\right]\left[ (k_v^r)^2+{2\omega\over \xi_r c_r}\right]^2
 (k_v^l)^2\over
 \big[  (k_v^r)^2-(k_d^l)^2\big]\big[  (k_v^r)^2-(k_d^r)^2\big]\left[  (k_v^l)^2+{2\omega\over \xi_l c_l}\right]^2\big[  {k_v^l}+ {k_v^r}\big]^2}
 \left| {(D_1^l)^2\over (D_1^r)^2}
 \right| \ .
\eea
In both cases unitarity relations are satisfied.

\newpage

\section{}

\noindent In this appendix we give the perturbative results for the construction of the ``in'' modes for spatial step-like discontinuities at $x=0$ for $v\neq 0$ ( in the subsonic-subsonic case). The details of how to construct them are given in subsection \ref{subconf} with the help of Fig.\ 1. For simplicity, we will only give explicitly the amplitudes of the propagating $u,v$ modes.

\vspace{0.4cm}
\noindent {\bf $\bullet$ Mode $u_{\omega,\phi}^{v,in}$}
\vspace{0.4cm}

\noindent By solving the system (\ref{modevin}) we find, for the propagating modes, at $O(z_l^2)$ (where $z_l=\xi_l \omega/c_l$)
\begin{eqnarray}
  A_v^l &=& \frac{2 \sqrt{c_l c_r}}{c_l+c_r}+i\frac{c_l^{3/2} \left(c_l-c_r\right) c_r \left(\sqrt{c_l^2-v^2}-\sqrt{c_r^2-v^2}\right) z_l  }{\left(c_l-v\right) \left(v-c_r\right) \left(c_l+c_r\right) \sqrt{\left(v^2-c_l^2\right) c_r \left(v^2-c_r^2\right)}}+\nonumber \\
  &-&\frac{z_l^2c_l^{5/2} \left(c_l-c_r\right){}^2}{8 \left(v-c_l\right){}^3 \left(v+c_l\right){}^2 \left(v-c_r\right){}^3 c_r^{3/2} \left(v+c_r\right){}^2 \left(c_l+c_r\right)} \left[-v^3 \left(v-c_r\right){}^3 \left(v+c_r\right){}^2+\right.\nonumber \\
 &+&\left. v^2 c_l \left(v-c_r\right){}^3 \left(v+c_r\right){}^2+c_l^5 \left(v^3-v^2 c_r+3 v c_r^2+c_r^3\right)+v c_l^4 \left(-v^3+c_r \left(v^2+c_r \left(v+3 c_r\right)\right)\right)+\right.\nonumber \\
  &+&\left.c_l^3 \left(-2 v^5+c_r \left(2 v^4+c_r \left(v+c_r\right) \left(-5 v^2+2 v c_r+c_r^2-4 \sqrt{\left(v^2-c_l^2\right) \left(v^2-c_r^2\right)}\right)\right)\right)+\right.\nonumber \\
  &+&\left.v c_l^2 \left(2 v^5+c_r \left(-2 v^4+c_r \left(v+c_r\right) \left(-3 v^2-2 v c_r+3 c_r^2-4 \sqrt{\left(v^2-c_l^2\right) \left(v^2-c_r^2\right)}\right)\right)\right)\right],\\\non
\eea

\bea
  A_u^r &=&\frac{c_l-c_r}{c_l+c_r}-i\frac{c_l^2 \left(c_l-c_r\right) c_r \left(v^2-c_r^2+\sqrt{\left(c_l^2-v^2\right) \left(c_r^2-v^2\right)}\right) z_l  }{\sqrt{c_l^2-v^2} \left(c_l+c_r\right) \left(v^2-c_r^2\right){}^2}+\nonumber \\
  &+&\frac{c_l^3 \left(c_l-c_r\right)z_l^2 }{4 \left(v^2-c_l^2\right){}^2 c_r \left(c_l+c_r\right) \left(v^2-c_r^2\right){}^3} \left[2 c_l^5 c_r^3-v^2 \left(v^2-c_r^2\right){}^3+c_l^4 \left(-v^4+c_r^4\right)\right.+\nonumber \\
  &+&\left.c_l^2 \left(2 v^6-v^4 c_r^2-2 v^2 c_r^4+c_r^6\right)+2 c_l^3 c_r^3 \left(-3 v^2+c_r^2-2 \sqrt{\left(v^2-c_l^2\right) \left(v^2-c_r^2\right)}\right)+\right.\nonumber \\
  &+&\left.2 v^2 c_l c_r^3 \left(-c_r^2+2 \left(v^2+\sqrt{\left(v^2-c_l^2\right) \left(v^2-c_r^2\right)}\right)\right)\right]
\end{eqnarray}
The important check is the unitarity relation $|A_v^l|^2+|A_u^r|^2=1$, which is satisfied quite non-trivially at $O(z_l^2)$, as
\begin{eqnarray}
  |A_v^l|^2 &=& \frac{4c_lc_r}{(c_l+c_r)^2}-\frac{ z_l^2c_l^3 \left(c_l-c_r\right){}^2}{2 \left(v-c_l\right){}^3 \left(v+c_l\right){}^2 \left(v-c_r\right){}^3 c_r \left(v+c_r\right){}^2 \left(c_l+c_r\right){}^2}  \left[v^3 \left(v-c_r\right){}^3 \left(v+c_r\right){}^2+\right.\nonumber\\
  &-&\left.v^2 c_l \left(v-c_r\right){}^3 \left(v+c_r\right){}^2+v c_l^4 \left(v-c_r\right) \left(v^2+c_r^2\right)-c_l^5 \left(v-c_r\right) \left(v^2+c_r^2\right)+\right.\nonumber\\
  &+&\left.c_l^2(c_l-v) \left(2 v^5+c_r \left(-2 v^4+c_r \left(v+c_r\right) \left(v-c_r\right)^2\right)\right)\right]\ ,\\
|A_u^r|^2 &=&\frac{c_l-c_r}{c_l+c_r}+\frac{c_l^2 \left(c_l-c_r\right) c_r \left(v^2-c_r^2+\sqrt{\left(v^2-c_l^2\right) \left(v^2-c_r^2\right)}\right) z_l }{\sqrt{v^2-c_l^2} \left(c_l+c_r\right) \left(v^2-c_r^2\right){}^2}+\nonumber \\
  &+&\frac{c_l^3 \left(c_l-c_r\right) z_l^2 }{4 \left(v^2-c_l^2\right){}^2 c_r \left(c_l+c_r\right) \left(v^2-c_r^2\right){}^3}\left[2 c_l^5 c_r^3-v^2 \left(v^2-c_r^2\right){}^3+c_l^4 \left(-v^4+c_r^4\right)+\right.\nonumber \\
  &+&\left.c_l^2 \left(2 v^6-v^4 c_r^2-2 v^2 c_r^4+c_r^6\right)+2 c_l^3 c_r^3 \left(-3 v^2+c_r^2-2 \sqrt{\left(v^2-c_l^2\right) \left(v^2-c_r^2\right)}\right)+\right.\nonumber \\
  &+&\left.2 v^2 c_l c_r^3 \left(-c_r^2+2 \left(v^2+\sqrt{\left(v^2-c_l^2\right) \left(v^2-c_r^2\right)}\right)\right)\right].
\end{eqnarray}

\newpage

\bigskip
\noindent {\bf $\bullet$ Modes $u_{\omega,\phi}^{uin}$}
\bigskip

\noindent The construction proceeds from Eq (\ref{modeuin}), which gives, at $O(z_l^2)$
\begin{eqnarray}\non
  A_v^l &=& \frac{c_r-c_l}{c_l+c_r}-i\frac{c_l^3 \left(c_l-c_r\right) \left(\sqrt{c_l^2-v^2}-\sqrt{c_r^2-v^2}\right) z_l  }{\left(c_l^2-v^2\right){}^{3/2} \left(c_l+c_r\right) \sqrt{c_r^2-v^2}}+\\\non
  &+&\frac{c_l^3 \left(c_l-c_r\right)z_l^2 }{4 \left(c_l^2-v^2\right){}^3 c_r \left(c_l+c_r\right) \left(v^2-c_r^2\right){}^2}\left[-v^4 \left(v^2-c_r^2\right){}^2+v^4 c_l^2 \left(3 v^2-c_r^2\right)+2 c_l^5 c_r \left(c_r^2-v^2\right)+\right.\nonumber \\
  &+&\left.c_l^6 \left(v^2+c_r^2\right)+c_l^4 \left(-3 v^4-2 v^2 c_r^2+c_r^4\right)+2 c_l^3 c_r \left(c_r^2-v^2\right) \left(c_r^2-2 \left(v^2+\sqrt{\left(v^2-c_l^2\right) \left(v^2-c_r^2\right)}\right)\right)\right],\\
A_u^r &=&\frac{2 \sqrt{c_l c_r}}{c_l+c_r}-i\frac{c_l^3\sqrt{c_r} \left(c_l-c_r\right) \left(\sqrt{c_l^2-v^2}-\sqrt{c_r^2-v^2}\right) z_l }{\left(v+c_l\right){}^{3/2} \left(v+c_r\right) \left(c_l+c_r\right) \sqrt{c_l \left(c_l-v\right) \left(c_r^2-v^2\right)}}+\nonumber \\
&+&\frac{c_l^{5/2} \left(c_l-c_r\right){}^2z_l^2 }{8 \left(v-c_l\right){}^2 \left(v+c_l\right){}^3 \left(v-c_r\right){}^2 c_r^{3/2} \left(v+c_r\right){}^3 \left(c_l+c_r\right)} \left[v^3 \left(v-c_r\right){}^2 \left(v+c_r\right){}^3+\right.\nonumber \\
&+&\left.v^2 c_l \left(v-c_r\right){}^2 \left(v+c_r\right){}^3+c_l^5 \left(v^3+c_r \left(v^2+3 v c_r-c_r^2\right)\right)+v c_l^4 \left(v^3+c_r \left(v^2-v c_r+3 c_r^2\right)\right)+\right.\nonumber \\
&-&\left.v c_l^2 \left(2 v^5+c_r \left(2 v^4-\left(v-c_r\right) c_r \left(3 v^2-2 v c_r-3 c_r^2+4 \sqrt{\left(v^2-c_l^2\right) \left(v^2-c_r^2\right)}\right)\right)\right)+\right.\nonumber \\
&-&\left.c_l^3 \left(2 v^5+c_r \left(2 v^4+\left(v-c_r\right) c_r \left(5 v^2+2 v c_r-c_r^2+4 \sqrt{\left(v^2-c_l^2\right) \left(v^2-c_r^2\right)}\right)\right)\right)\right].
\end{eqnarray}
The unitarity relation $|A_v^l|^2+|A_u^r|^2=1$ is again non trivially satisfied at $O(z_l^2)$, being
\begin{eqnarray}
  |A_v^l|^2 &=& \left(\frac{c_r-c_l}{c_l+c_r}\right)^2-
  \frac{z_l^2 c_l^3 \left(c_l-c_r\right){}^2}{2 \left(v^2-c_l^2\right){}^3 c_r \left(c_l+c_r\right){}^2 \left(v^2-c_r^2\right){}^2}\left[v^4 \left(v^2-c_r^2\right){}^2+v^4 c_l^2 \left(c_r^2-3 v^2\right)+\right.\nonumber \\
  &-&\left.c_l^6 \left(v^2+c_r^2\right)+c_l^4 \left(3 v^4+2 v^2 c_r^2-c_r^4\right)\right],\\\non
|A_u^r|^2 &=&\frac{4c_lc_r}{(c_l+c_r)^2}+\frac{z_l^2 c_l^3 \left(c_l-c_r\right){}^2 }{2 \left(v-c_l\right){}^2 \left(v+c_l\right){}^3 \left(v-c_r\right){}^2 c_r \left(v+c_r\right){}^3 \left(c_l+c_r\right){}^2}\left[v^3 \left(v-c_r\right){}^2 \left(v+c_r\right){}^3+\right.\nonumber\\
  &+&\left.v^2 c_l \left(v-c_r\right){}^2 \left(v+c_r\right){}^3+v c_l^4 \left(v+c_r\right) \left(v^2+c_r^2\right)+c_l^5 \left(v+c_r\right) \left(v^2+c_r^2\right)+\right.\nonumber\\
  &+&\left.c_l^2(c_l+v) \left(-2 v^5+c_r \left(-2 v^4-\left(v-c_r\right) c_r \left(v+c_r\right)^2\right)\right)\right].
\end{eqnarray}

\section{}

\noindent In this appendix we extend the leading order results of the calculations of the amplitudes of the propagating modes  $u_{\omega,\phi}^{3,in}$ and $u_{\omega,\phi}^{4,in*}$, in the case of the subsonic-supersonic spatial step-like discontinuity. With these, we are able  to check the unitarity relations.
%
%\begin{itemize}
%\item $u_{\omega}^{1in}$ mode
%\end{itemize}
%
%The amplitudes regarding Fig \ref{fig:supsub1in} read:
%\begin{eqnarray*}
%% \nonumber to remove numbering (before each equation)
%  A_u^r &=& \frac{v+c_r}{v-c_r} \\
%  A_v^l &=& \sqrt{\frac{c_r}{c_l}}\frac{v-c_l}{v-c_r} \\
%  A_u^l &=& \sqrt{\frac{c_r}{c_l}}\frac{v+c_l}{c_r-v}
%\end{eqnarray*}
%
%where
%\begin{eqnarray*}
%% \nonumber to remove numbering (before each equation)
%  |A_u^r|^2 &=& \frac{(v+c_r)^2}{(v-c_r)^2} \\
%  |A_v^l|^2 &=& \frac{c_r(v-c_l)^2}{c_l(v-c_r)^2} \\
%  |A_u^l|^2 &=& \frac{c_r(v+c_l)^2}{c_l(c_r-v)^2}
%\end{eqnarray*}
%
%These amplitudes satisfy $|A_2^r|^2+|A_1^l|^2-|A_2^l|^2=1$.\\

\bigskip
\noindent {\bf $\bullet$ Modes $u_{\omega,\phi}^{4in*}$}
\bigskip

\noindent The amplitudes depicted in Fig.\  4 are:
\begin{eqnarray}
  A_u^{r'}& = &\frac{\sqrt{2c_r}(v^2-c_l^2)^{3/4}(v+c_r)}{c_l\sqrt{z_l}(c_r^2-c_l^2)\sqrt{c_r^2-v^2}}\left(\sqrt{c_r^2-v^2}-i\sqrt{v^2-c_l^2}\right)+(A+iB)\sqrt{z_l}\ ,\\
  A_v^{l'} &=& \frac{(v^2-c_l^2)^{3/4}(v+c_r)}{c_l^{3/2}\sqrt{2z_l}(c_l+c_r)\sqrt{c_r^2-v^2}}\left(\sqrt{c_r^2-v^2}-i\sqrt{v^2-c_l^2}\right)+(C+iD)\sqrt{z_l}\ , \\
  A_u^{l'} &=& \frac{(v^2-c_l^2)^{3/4}(v+c_r)}{c_l^{3/2}\sqrt{2z_l}(c_l-c_r)\sqrt{c_r^2-v^2}}\left(\sqrt{c_r^2-v^2}-i\sqrt{v^2-c_l^2}\right)+(E+iF)\sqrt{z_l}\ .
\end{eqnarray}
where
\bea
  A &=&\frac{\left(-8 v^6+2 v c_l^4 \left(-2 v+c_r\right)+2 v^2 c_r \left(v^3+4 v^2 c_r-2 c_r^3\right)+c_l^2 \left(7 v^4-c_r \left(4 v^3-2 v^2 c_r+c_r^3\right)\right)\right) c_l\sqrt{   c_r z_l}}{2 \sqrt{2} v \left(v^2-c_l^2\right){}^{3/4} \left(v-c_r\right){}^2 \left(v+c_r\right) \left(c_l^2-c_r^2\right)}\ ,\\
  B &=& -\frac{\left(8 v^3+6 v^2 c_r+c_l^2 \left(-v+c_r\right)\right) c_l\sqrt{   c_r \left(-v^2+c_r^2\right) z_l}}{2 \sqrt{2} v \left(v^2-c_l^2\right){}^{1/4} \left(v^2-c_r^2\right) \left(c_l^2-c_r^2\right)}\ ,\\
 C &=&\frac{1}{4\sqrt{2}v(v^2-c_l^2)^{3/4}(v-c_r)^2(v+c_r) (c_l+c_r)}\Big[8 v^6+2 v^2 c_l^4+2 v c_l c_r^2(v^2-c_r^2)+2 v c_l^3(-v^2+c_r^2) +\\\non
 &+& c_l^2(-7 v^4+2 v^3 c_r+2 v^2 c_r^2-2 v c_r^3+c_r^4)+2 v^2 c_r \left(-v^3+c_r \left(-4 v^2+c_r(v+c_r)\right)\right)\Big]\sqrt{c_lz_l}\ ,\\
 D &=&-\frac{ \left(8 v^3+6 v^2 c_r+c_l^2 \left(-v+c_r\right)\right) \sqrt{c_lz _l}}{4 \sqrt{2} v \left(v^2-c_l^2\right){}^{1/4} \left(c_l+c_r\right) \sqrt{ \left(-v^2+c_r^2\right) }}\ ,\\
E &=&\frac{1}{4 \sqrt{2} v \left(v^2-c_l^2\right){}^{3/4} \left(v-c_r\right){}^2 \left(c_l-c_r\right) \left(v+c_r\right)}\Big[8 v^6+v^2 c_l^2 \left(-7 v^2+2 c_l \left(v+c_l\right)\right)+\\\non
&+&2 v^3 \left(-v^2+c_l^2\right) c_r-2 v \left(4 v^3+c_l \left(v^2-v c_l+c_l^2\right)\right) c_r^2+2 v \left(v^2-c_l^2\right) c_r^3+\left(2 v^2+2 v c_l+c_l^2\right) c_r^4\Big] \sqrt{ c_lz_l}\ ,\\
F &=&-\frac{ \left(8 v^3+6 v^2 c_r+c_l^2 \left(-v+c_r\right)\right) \sqrt{c_lz_l}}{4 \sqrt{2} v \left(v^2-c_l^2\right){}^{1/4} \left(c_l-c_r\right) \sqrt{\left(-v^2+c_r^2\right)}}\ .
\eea
Their squared modulus read
\bea
  |A_2^{r'}|^2 &=& \frac{2c_r(v^2-c_l^2)^{3/2}(v+c_r)}{c_l^2z_l(c_l^2-c_r^2)(v-c_r)}+G\ ,\\
  |A_1^{l'}|^2 &=& \frac{(v^2-c_l^2)^{3/2}(v+c_r)(c_r-c_l)}{2c_l^3z_l(c_l+c_r)(c_r-v)}+ H\ ,\\
  |(A_2^l)^{\Omega<0}|^2 &=& \frac{(v^2-c_l^2)^{3/2}(v+c_r)(c_l+c_r)}{2c_l^3z_l(c_l-c_r)(v-c_r)}+I\ ,
\eea
where
\bea
  G &=& \frac{c_r \left(-2 v^4+4 v^2 c_r^2+c_l^2 \left(-3 v^2+c_r^2\right)\right)}{v \left(v-c_r\right){}^2 \left(-c_l^2+c_r^2\right)}\ ,  \\
  H &=& \frac{\left(c_l-c_r\right) \left(2 v^4-2 v^2 c_r \left(v+c_r\right)+c_l^2 \left(v^2+2 v c_r-c_r^2\right)+2 v c_l \left(-v^2+c_r^2\right)\right)}{4 v c_l \left(v-c_r\right){}^2 \left(c_l+c_r\right)}\ , \\
  I &=& \frac{\left(c_l+c_r\right) \left(2 v^4-2 v^2 c_r \left(v+c_r\right)+2 v c_l \left(v^2-c_r^2\right)+c_l^2 \left(v^2+2 v c_r-c_r^2\right)\right)}{4 v c_l \left(v-c_r\right){}^2 \left(c_l-c_r\right)}\ .
\eea
The above amplitudes satisfy the unitarity condition $|A_u^{r'}|^2+|A_v^{l'}|^2-|A_u^{l'}|^2=-1$ at this perturbative level.

\bigskip
\noindent {\bf $\bullet$ Modes $u_{\omega,\phi}^{3,in}$}
\bigskip

\noindent The amplitudes sketched in Fig.\  \ref{fig:sub_sup} turn out to be
\bea
 A_u^R &=&\frac{\sqrt{2c_r}(v^2-c_l^2)^{3/4}(v+c_r)}{c_l\sqrt{z_l}(c_r^2-c_l^2)\sqrt{c_r^2-v^2}}\left(\sqrt{c_r^2-v^2}+i\sqrt{v^2-c_l^2}\right)-(A-iB)\sqrt{z_l}\ ,  \\
  A_v^L &=&\frac{(v^2-c_l^2)^{3/4}(v+c_r)}{c_l^{3/2}\sqrt{2z_l}(c_l+c_r)\sqrt{c_r^2-v^2}}\left(\sqrt{c_r^2-v^2}+i\sqrt{v^2-c_l^2}\right)-(C-D)\sqrt{z_l}\ , \\
  A_u^L &=&\frac{(v^2-c_l^2)^{3/4}(v+c_r)}{c_l^{3/2}\sqrt{2z_l}(c_l-c_r)\sqrt{c_r^2-v^2}}\left(\sqrt{c_r^2-v^2}+i\sqrt{v^2-c_l^2}\right)-(E-iF)\sqrt{z_l}\ ,
\eea
where
\bea
  &&|A_u^R|^2 = \frac{2c_r(v^2-c_l^2)^{3/2}(v+c_r)}{c_l^2z_l(c_l^2-c_r^2)(v-c_r)}-G\ ,\\
  &&|A_v^L|^2 = \frac{(v^2-c_l^2)^{3/2}(v+c_r)(c_r-c_l)}{2c_l^3z_l(c_l+c_r)(c_r-v)}- H\ ,\\
  &&|A_u^L|^2 = \frac{(v^2-c_l^2)^{3/2}(v+c_r)(c_l+c_r)}{2c_l^3z_l(c_l-c_r)(v-c_r)}-I\ .
\eea
Again, one sees that $|A_u^R|^2+|A_v^L|^2-|A_u^L|^2=1$ is satisfied.

\end{document}